%% file: main.tex
\documentclass{article}

\usepackage{authblk}

\usepackage{amsmath,amssymb}
\usepackage{mathrsfs}
\usepackage[numbers,sort&compress]{natbib}

\usepackage{hyperref}
\usepackage{bbm}
\usepackage[paper=a4paper,text={17cm,24cm},centering]{geometry}
\usepackage{subcaption}
\usepackage{enumerate}
\usepackage{dsfont}
\usepackage{physics}
\usepackage{mdframed}
\usepackage{adjustbox}
\usepackage{makecell}

\usepackage{tikz}
\usetikzlibrary{arrows.meta,arrows}
\usepackage[compat=1.1.0]{tikz-feynman}

\usepackage{titlesec}
\titleformat{\paragraph}[block]{\normalfont\normalsize\bfseries}{\theparagraph}{1em}{}

\def\be{\begin{equation}}
\def\ee{\end{equation}}
\def\MSbar{{\overline{\text{MS}}}}
\input graphics.tikz

\title{Unitary perturbation theory on the light cone\\
using adiabatic switching}
\author{Stéphane Munier}
\affil{\it\normalsize CPHT, CNRS, École polytechnique, Institut Polytechnique de Paris, 91120 Palaiseau, France}
\date{October 8, 2025; v2: November 20, 2025}

\begin{document}

\maketitle

\begin{abstract}
Light-cone perturbation theory is a powerful tool for calculating high-energy scattering amplitudes, particularly for quantum particles such as electrons, photons, or protons scattering off heavy nuclei, a process analogous to potential scattering. Central to these computations are the light-cone wave functions of incoming and outgoing particles, representing the projection of dressed initial and final states onto partonic Fock states. The dressed states are obtained by applying an evolution operator in the Dirac picture to bare partonic states, which may be interpreted physically as a time evolution from preparation to interaction. In standard approaches, a non-unitary operator is used, and proper normalization is imposed a posteriori. Here, we systematically develop perturbation theory from a perturbatively unitary evolution operator, using adiabatic switching to regularize the infinite-time limits. This provides a theoretically coherent framework for organizing calculations, reproducing known results entirely diagrammatically without enforcing unitarity by hand. We illustrate the method with a simple quantum mechanical model, enabling calculations to arbitrary perturbative orders, and then evaluate wave functions in field theories quantized on the light cone, focusing on a massive scalar theory with cubic interaction at one-loop accuracy.
\end{abstract}


\section{Introduction}

Light-cone quantization has been an active area of research since the late 1960s \cite{Weinberg:1966jm,Kogut:1969xa,Bjorken:1970ah,Lepage:1980fj,Pauli:1985pv}, largely motivated by the quest for a non-perturbative understanding of hadrons. The development of quantum chromodynamics (QCD) at high energies, particularly from the 1990s onward with experiments such as DESY-HERA and CERN-LHC, renewed interest in the calculation of wave functions in the framework of light-cone quantization. This is essentially because the scattering of an electron, photon or proton off a large nucleus at very high energy ``looks like'' the scattering of a quantum particle moving on the light cone off a localized potential, or ``shockwave'', when viewed in the rest frame of the nucleus. In this context, it is natural to describe the incoming particle in terms of its partonic wave functions, making a non-covariant, time-ordered formulation of perturbation theory in a light-cone frame particularly well-suited. A wealth of calculations at next-to-leading order accuracy is now available for various processes relevant to phenomenology, especially at CERN-LHC and the future electron-ion collider (EIC); see, e.g., Refs.~\cite{Beuf:2016wdz,Lappi:2016oup,Beuf:2017bpd,Taels:2022tza,Taels:2023czt,Beuf:2024msh}.

Such a formulation is not only natural from a physical perspective, but may also be technically advantageous. A prominent example is the rederivation of the leading-order Balitsky-Fadin-Kuraev-Lipatov (BFKL) equation \cite{Kuraev:1977fs,Balitsky:1978ic} in the color dipole model~\cite{Mueller:1993rr}, based on the calculation of light-cone wave functions of an onium. The color dipole model turns out to naturally reproduce the BFKL equation, basically from a simple tree-level diagram, appropriately iterated.\footnote{The color dipole model assumes the limit of an infinite number of colors, and strictly speaking, the derivation of the BFKL equation in this framework is fully rigorous only in that limit. Nevertheless, the resulting equation coincides with the original one.} Interestingly enough, it also enables the study of unitarity corrections~\cite{Salam:1995uy,Kovchegov:1999yj}, which are much harder to implement in other approaches.\footnote{See e.g. Refs.~\cite{Bartels:1992ym,Bartels:1999aw}.} We refer the reader to our recent review~\cite{Angelopoulou:2023qdm} for an elementary introduction to light-cone perturbation theory (LCPT) and the dipole model, from which we adopt most notations and conventions. For further background on high-energy scattering from a modern perspective, see e.g. the textbook \cite{Kovchegov:2012mbw}.

The calculation of wave functions in LCPT requires a prescription to define the  asymptotic limits of the time-evolution operator, whose matrix elements between states of fixed momentum essentially yield the wave functions we need. The vast majority of LCPT calculations in the literature adopt the prescription physically motivated in the landmark paper by Gell-Mann and Goldberger~\cite{Gell-Mann:1953dcn}, later adapted to the light-cone framework in foundational works such as Refs.~\cite{Bjorken:1970ah,Lepage:1980fj}. We shall refer to this formulation as the \textit{standard prescription}. An alternative is the so-called \textit{adiabatic switching prescription}. Although used less frequently, it has played a central role in key developments, notably in the proof of the Gell-Mann--Low theorem~\cite{Gell-Mann:1951ooy}, and in high-energy QCD it was instrumental for the development of the color dipole model.\footnote{See Appendix~A of Ref.~\cite{Chen:1995pa}, where a rigorous implementation of the adiabatic switching prescription is essential. For completeness, we note that in the original paper~\cite{Mueller:1993rr}, the precise formulation of this prescription  was immaterial, since the only field-theoretical calculation required was that of real radiation at leading order, and the virtual corrections could be fully obtained via unitarity. In contrast, Ref.~\cite{Chen:1995pa} computed these corrections explicitly from the diagrams, with the aim of verifying the original derivation in detail.}

A major advantage of the adiabatic switching framework is that it yields a unitary time-evolution operator.\footnote{Recall that our discussion is restricted to perturbation theory.} Properly normalized wave functions then follow from a purely diagrammatic expansion. In contrast, in the standard formulation, overall normalization is not computed diagrammatically; unitarity must be imposed by hand.\footnote{For a practical example, see Appendix~E of Ref.~\cite{Taels:2023czt}.}

In practice, the distinction between these two prescriptions is often not clearly stated in the literature.\footnote{For instance, in Ref.~\cite{Zhang:1993dd}, the perturbative formula given in Eq.~(3.26) does not follow directly from the adiabatic prescription defined in Eq.~(3.14), even though one might expect it to.
We stress, however, that this does not call into question the results obtained using Eq.~(3.26), since the latter is correct within the standard prescription.} In this work, we revisit the early formulations of time-ordered perturbation theory in both the standard and adiabatic switching prescriptions. Our main objective is to clarify potential ambiguities and to put this comparison to work through fully explicit computations, in setups simple enough to be tractable yet sufficiently rich that careful control is essential. Beyond its potential pedagogical value, our analysis provides a framework that can be extended to phenomenologically relevant settings.

The remainder of this paper is organized as follows. Section~\ref{sec:PT} is devoted to the construction of perturbation theory, developed in parallel using both prescriptions. In Section~\ref{sec:0D}, we illustrate the formalism by computing arbitrary wave functions in the simplest quantum mechanical model. Section~\ref{sec:phi3} turns to the simplest relevant quantum field theory, namely scalar theory with cubic interaction. There, we compute at one-loop accuracy the probability amplitude to find a single asymptotic particle in either a one- or two-particle Fock state. Section~\ref{sec:conclusion} summarizes our findings and perspectives. An Appendix collects useful formulas.


\section{Perturbation theory}
\label{sec:PT}

This section reviews two formulations of time-ordered perturbation theory in quantum field theory, which differ in how the evolution operator is regularized in the infinite-time limit. After specifying the setting, we first present the standard formulation, as laid out in the classical reference~\cite{Gell-Mann:1953dcn}, and then turn to the adiabatic switching prescription, discussed for example in the early review~\cite{DeWitt:1956jua} and in textbooks such as Ref.~\cite{Fetter:2003}.


\subsection{Theoretical setup}

We consider a theory defined by a Hamiltonian \( H \), which we assume can be decomposed additively into a free part \( H_0 \) and an interaction part \( H_1 \). The interaction Hamiltonian \( H_1 \) is characterized by the fact that it vanishes when the coupling constant \( \lambda \), assumed to be unique, is set to zero.

In this section, we assume that both $H$ and $H_0$ admit countable, complete, orthonormal sets of eigenvectors that form bases of the Hilbert space, with all eigenvalues non-degenerate. We denote these sets by $\{ \ket{\Psi_j} \}$ and $\{ \ket{\Phi_j} \}$, respectively. The corresponding eigenvalues are written as $\{ E_j \}$ and $\{ \varepsilon_j \}$:
\begin{equation}
H \ket{\Psi_j} = E_j \ket{\Psi_j}, 
\qquad 
H_0 \ket{\Phi_j} = \varepsilon_j \ket{\Phi_j}.
\label{eq:eigenstates-H}
\end{equation}
We shall regard the \(\ket{\Phi_i}\)'s as (appropriately symmetrized and normalized) tensor products of single-particle states,  and we may occasionally refer to them as ``Fock states,'' as a shorthand for the vectors of the canonical basis of the Fock space.
We further assume that the eigenstates of $H$ reduce continuously to those of $H_0$ in the limit where the interaction is turned off, in the sense that
\begin{equation}
\ket{\Psi_j} \xrightarrow[H_1 \to 0]{} \ket{\Phi_j}.
\end{equation}
We will refer to $\ket{\Psi_j}$ as the ``dressed state'' corresponding to the ``bare state'' $\ket{\Phi_j}$.
All these assumptions are not trivial and, in general, are not satisfied in standard quantum field theories such as scalar \( \varphi^3 \) theory, quantum electrodynamics (QED), or quantum chromodynamics (QCD). Nevertheless, it is sufficient for our purpose to develop the general formalism in an idealized setting. The additional complications that arise when these assumptions fail will be addressed in the specific examples considered later.

We will frequently make use of the decomposition of the identity operator in terms of projectors onto the eigenspaces of \( H_0 \):
\be
\mathbbm{I} = \sum_j \ket{\Phi_j} \bra{\Phi_j}.
\label{eq:completeness}
\ee

We will find it convenient to shift the spectrum of \( H_0 \), replacing the eigenvalues \( \varepsilon_j \) by new values \( \varepsilon_{jR} \), to be specified as needed. This modification leaves the physical content of the theory unchanged and can be implemented by adding to $H_0$ and subtracting from $H_1$ a diagonal operator:
\be
\Delta_R \equiv \sum_j (\varepsilon_{jR} - \varepsilon_j) \ket{\Phi_j} \bra{\Phi_j},
\ee
which amounts to the redefinitions
\be
H_{0} \rightarrow \mathcal{H}_{0R} \equiv H_{0} + \Delta_R, \quad 
H_{1} \rightarrow \mathcal{H}_{1R} \equiv H_{1} - \Delta_R.
\label{eq:split-Hamiltonian}
\ee
When the $\varepsilon_{jR}$'s are set at the eigenvalues $E_j$ of $H$, then we will just drop the subscript $R$: ${\cal H}_{0R}\to{\cal H}_0$, ${\cal H}_{1R}\to {\cal H}_1$.

Finally, we anticipate that the following discussion will apply directly to the quantum mechanical problem examined in the next section~\ref{sec:0D}. Through the introduction of the renormalized energies $\varepsilon_{jR}$, this discussion can be extended to simple quantum field theories, such as the massive scalar field (see Sec.~\ref{sec:phi3}). However, for other theories, even as simple as the massless scalar field, some of our assumptions would become problematic.


\subsection{Standard prescription}

\subsubsection{Formulation}

The time-dependence of a Schrödinger-picture state $\ket{\Psi(t)}_S$ is governed by the Schrödinger equation 
\be
i\frac{d}{dt}\ket{\Psi(t)}_S=H\ket{\Psi(t)}_S,
\label{eq:basic-schrodinger}
\ee
whose solution can be encoded in the evolution operator
\be
U_S(t,t_0)\equiv e^{-i(t-t_0)H}:\quad \ket{\Psi(t)}_S=U_S(t,t_0)\ket{\Psi(t_0)}_S.
\label{eq:U-from-H}
\ee
Obviously, the operator $U_S(t,t_0)$ itself obeys the Schrödinger equation~(\ref{eq:basic-schrodinger}) with the initial condition 
\be
U_S(t=t_0,t_0)={\mathbb I}.
\ee

Perturbation theory is most conveniently formulated in the Dirac, or interaction, picture. The states are redefined with respect to those in the Schrödinger picture as $\ket{\Psi(t)}_{IR}\equiv e^{it{\cal H}_{0R}}\ket{\Psi(t)}_S$, in such a way that $\ket{\Psi(0)}_{IR}=\ket{\Psi(0)}_S$. We can define the equivalent Heisenberg-picture time-independent state $\ket{\Psi}$ as the one that coincides with the latter vectors, thus justifying the substitution $\ket{\Psi(0)}_{IR}=\ket{\Psi}$ we shall make use of below. 

The corresponding interaction-picture evolution operator, that maps \(\ket{\Psi(t_0)}_{IR}\) to \(\ket{\Psi(t)}_{IR}\), obviously reads
\be
U_{IR}(t,t_0)=e^{it{\cal H}_{0R}}e^{-i(t-t_0)H}e^{-it_0{\cal H}_{0R}}.
\label{eq:interaction-picture-schrodinger-0}
\ee
For our purposes, a particularly useful form for developing perturbation theory is most easily derived by beginning with the Schrödinger equation expressed in the interaction picture,
\be
i\frac{\partial}{\partial t}{U_{IR}(t,t_0)}={\mathcal H}_{IR}(t){U_{IR}(t,t_0)},
\quad\text{with}\quad
{\mathcal H}_{IR}(t)\equiv e^{it{\mathcal H}_{0R}}{\mathcal H}_{1R} e^{-it{\mathcal H}_{0R}},
\label{eq:interaction-picture-schrodinger}
\ee
which we solve iteratively given the initial condition $U_{IR}(t=t_0,t_0)={\mathbbm I}$. Because of the time-dependence of the Dirac-picture Hamiltonian, instead of the simple exponential~(\ref{eq:U-from-H}), we get the following Dyson series:
\be
U_{IR}(t,t_0)={\mathbbm I}+\sum_{n=1}^\infty (-i)^n\int_{t_0}^t dt_1\cdots dt_n {\mathbbm 1}_{\{t_1\geq\cdots\geq t_{n}\}}{\mathcal H}_{IR}(t_1)\cdots{\mathcal H}_{IR}(t_n)
\equiv T\exp\left(-i\int_{t_0}^{t}dt' {\mathcal H}_{IR}(t')\right).
\label{eq:evolution-op-standard}
\ee
The states and operators in the interaction picture depend on how the Hamiltonian is split, see Eq.~(\ref{eq:split-Hamiltonian}). We shall omit the ``$R$'' subscripts on the interaction-picture states and operators when the energies $\varepsilon_{jR}$ are chosen to coincide with $E_j$. This will be the case throughout this section.

Following Gell-Mann and Goldberger in Ref.~\cite{Gell-Mann:1953dcn}, we consider the system to be prepared at some initial time $t_0<0$ in an eigenstate $\ket{\Phi_i}$ of ${\mathcal H}_0$. It is then evolved up to a later time $t>t_0$ (with $t\leq 0$, by convention), at which it is ``observed,'' or interacts. Technically, the initial state vector is multiplied by the evolution operator~(\ref{eq:evolution-op-standard}). In practice, however, the preparation time $t_0$ is not precisely known, so it is natural to average over it. The simplest choice is to introduce an exponential weight with parameter $\epsilon$, a real positive number. This yields an expression for the evolved state in the form of a Laplace transform with respect to the starting time:
\be
\ket{\Psi_i^{\text{std}\,\epsilon}(t)}_I\equiv\epsilon\int_{-\infty}^0dt_0\, e^{\epsilon t_0}\,U_I(t,t_0)\ket{\Phi_i}.
\label{eq:Gell-Mann-averaging}
\ee
Eventually, one takes the limit $\epsilon \to 0$, which pushes the typical values of $t_0$ contributing to the integral toward infinitely negative times, consistent with the physical expectation that the preparation time should be macroscopic and much longer than the typical duration of the interaction of this state with another state or with an external field.

Let us restrict our discussion to the state vector at time $t=0$, in such a way that the left-hand side of Eq.~(\ref{eq:Gell-Mann-averaging}) is the Heisenberg state $\ket{\Psi_i^{\text{std}\,\epsilon}}$. We insert the definition~(\ref{eq:evolution-op-standard}) of the evolution operator into this equation, and see how the calculation works for a generic term in the Dyson series:
\be
\begin{aligned}
\epsilon\int_{-\infty}^0dt_0\, e^{\epsilon t_0} & (-i)^n \int_{t_0}^0 dt_1{\mathcal H}_I(t_1)\cdots\int_{t_0}^{t_{n-1}}dt_n  {\mathcal H}_I(t_n)\ket{\Phi_i}\\
=& (-i)^n\int_{-\infty}^0 dt_1{\mathcal H}_I(t_1)\cdots \int_{-\infty}^{t_{n-1}}dt_n{\mathcal H}_I(t_n)\left(\epsilon\int_{-\infty}^{t_n} dt_0\, e^{\epsilon t_0}\right)\ket{\Phi_i}\\
=& (-i)^n\int_{-\infty}^0 dt_1{\mathcal H}_I(t_1) \cdots\int_{-\infty}^{t_{n-1}}dt_n{\mathcal H}_I(t_n)\,e^{\epsilon t_n}\ket{\Phi_i}
\end{aligned}
\ee
Inserting the completeness relation~(\ref{eq:completeness}) before each factor ${\mathcal H}_I$ and using
\be
\mel{\Phi_{k_2}}{{\mathcal H}_I(t)}{\Phi_{k_1}}=e^{-i(E_{k_1}-E_{k_2})t}\mel{\Phi_{k_2}}{{\mathcal H}_1}{\Phi_{k_1}},
\ee
we can perform the time integrations. We get
\begin{multline}
{\ket{\Psi_i^{\text{std}\,\epsilon}}}=\ket{\Phi_i}+\sum_{j}\ket{\Phi_j}\frac{1}{E_i-E_j+i\epsilon}\mel{\Phi_j}{{\mathcal H}_1}{\Phi_i}\\
+\sum_{j,j_1}\ket{\Phi_{j}}\frac{1}{E_i-E_{j}+i\epsilon}\mel{\Phi_{j}}{{\mathcal H}_1}{\Phi_{j_1}}\frac{1}{E_i-E_{j_1}+i\epsilon}\mel{\Phi_{j_1}}{{\mathcal H}_1}{\Phi_i}\\
+\sum_{j,j_2,j_1}\ket{\Phi_{j}}\frac{1}{E_i-E_{j}+i\epsilon}\mel{\Phi_{j}}{{\mathcal H}_1}{\Phi_{j_2}}\frac{1}{E_i-E_{j_2}+i\epsilon}\mel{\Phi_{j_2}}{{\mathcal H}_1}{\Phi_{j_1}}\frac{1}{E_i-E_{j_1}+i\epsilon}\mel{\Phi_{j_1}}{{\mathcal H}_1}{\Phi_i}+\cdots
\label{eq:psi_i_eps-standard}
\end{multline}
Note that the same ``$i\epsilon$'' term appears in each energy denominator. We emphasize from the outset that the precise coefficients of $i\epsilon$ are important in such expressions, as these equations are not to be interpreted in the distributional sense in the present work: here, {\it $\epsilon$ is understood, \textit{a priori}, as a finite real positive parameter.}


\subsubsection{Constructing eigenstates of $H$ from the perturbative series}
\label{subsec:eigenstates-H}

From now on, we shall regard the particular asymptotic state $\ket{\Phi_i}$, whose dressed counterpart we are constructing, as representing a single-particle state.

The $\epsilon\to 0$ limit of Eq.~(\ref{eq:psi_i_eps-standard}) is singular because some energy denominators may vanish: $E_i-E_j=0$ if $i$ and $j$ label the same state. But we can reorder the terms in this expansion and eventually factorize the singular overlap ${\braket{\Phi_i}{\Psi_i^{\text{std}\,\epsilon}}}$ from sums over states that are different with respect to the asymptotic state. Then, within our assumptions, the $\epsilon\to 0$ limit of the ratio of $\ket{\Psi_i^{\text{std}\,\epsilon}}$ to its projection on an eigenstate of ${\cal H}_0$, e.g. ${\braket{\Phi_i}{\Psi_i^{\text{std}\,\epsilon}}}$ is regular. Namely, introducing
\be
\ket{\widetilde{\Psi_i^{\text{std}\,\epsilon}}}\equiv\frac{\ket{\Psi_i^{\text{std}\,\epsilon}}}{\braket{\Phi_i}{\Psi_i^{\text{std}\,\epsilon}}},
\quad\text{the limit state}\quad
\ket{\widetilde{\Psi_i^{\text{std}}}}\equiv\lim_{\epsilon\to 0}\ket{\widetilde{\Psi_i^{\text{std}\,\epsilon}}}
\ee
is well defined. Its expansion in terms of eigenstates of the free Hamiltonian reads
\begin{mdframed}[innertopmargin=-5pt, innerbottommargin=5pt, leftmargin=0pt, rightmargin=0pt,skipabove=2pt,skipbelow=2pt]
\begin{multline}
\ket{\widetilde{\Psi_i^{\text{std}}}}=
\ket{\Phi_i}+\sum_{j\neq i}\ket{\Phi_j}\Bigg(\frac{1}{E_i-E_j}\mel{\Phi_j}{{\mathcal H}_1}{\Phi_i}
+\sum_{j_1\neq i}\frac{1}{E_i-E_j}\mel{\Phi_{j}}{{\mathcal H}_1}{\Phi_{j_1}}\frac{1}{E_i-E_{j_1}}\mel{\Phi_{j_1}}{{\mathcal H}_1}{\Phi_i}\\
+\sum_{j_2,j_1\neq i}\frac{1}{E_i-E_j}\mel{\Phi_{j}}{{\mathcal H}_1}{\Phi_{j_2}}\frac{1}{E_i-E_{j_2}}\mel{\Phi_{j_2}}{{\mathcal H}_1}{\Phi_{j_1}}\frac{1}{E_i-E_{j_1}}\mel{\Phi_{j_1}}{{\mathcal H}_1}{\Phi_i}+\cdots\Bigg)
\label{eq:standard-PT}
\end{multline}
\end{mdframed}
This is the standard formulation of perturbation theory. Such a formula was used in Ref.~\cite{Bjorken:1970ah} and in almost all modern calculations.

We may check that this well-defined state is an eigenstate of $H$ with eigenvalue $E_i$ by algebraic manipulations. To this aim, we rewrite the previous formula in a recursive form:
\be
\ket{\widetilde{\Psi_i^{\text{std}}}}=
\ket{\Phi_i}+\sum_{j\neq i}\ket{\Phi_j}\frac{1}{E_i-E_j}\mel{\Phi_j}{{\mathcal H}_1}{\widetilde{\Psi_i^{\text{std}}}}
\ee
on which it is easy to act $H$. A few straightforward manipulations enable to see explicitly that, indeed, $H\ket{\widetilde{\Psi_i^{\text{std}}}}=E_i\ket{\widetilde{\Psi_i^{\text{std}}}}$. The state $\ket{\widetilde{\Psi_i^{\text{std}}}}$ obeys the equation~(\ref{eq:eigenstates-H}) that defines $\ket{\Psi_i}$, and is thus proportional to the latter, within our assumptions of non-degeneracy.

The energies $E_i$ can be determined analytically from the bare energies $\varepsilon_{iR}$, that appear in the original Hamiltonian, order-by-order in perturbation theory. To this aim, we observe that
\be
\mel{\Phi_i}{{H}_{1}}{\widetilde{\Psi_i^{\text{std}}}}=\mel{\Phi_i}{(H-{H}_{0})}{\widetilde{\Psi_i^{\text{std}}}}=(E_i-\varepsilon_{i})\braket{\Phi_i}{\widetilde{\Psi_i^{\text{std}}}}\,,
\ee
from which we deduce the following identity:
\be
E_i-\varepsilon_{i}=\mel{\Phi_i}{{H}_{1}}{\widetilde{\Psi_i^{\text{std}}}}\,.
\label{eq:energy-shift}
\ee
Expanding this equation, we get the standard formula for the energy shift as obtained from time-independent perturbation theory:\footnote{See e.g. Eq.~(3.32) in Ref.~\cite{Zhang:1993dd}.}
\be
E_i-\varepsilon_i=\mel{\Phi_i}{H_1}{\Phi_i}+\sum_{j\neq i}\frac{\left|\mel{\Phi_j}{H_1}{\Phi_i}\right|^2}{E_i-E_j}+\cdots.
\ee


\subsubsection{Fock state expansion for normalized states}

We have organized Eq.~(\ref{eq:standard-PT}) as a weighted sum of the states \(\ket{\Phi_j}\). We denote the coefficient of \(\ket{\Phi_j}\) in this series, namely $\braket{\Phi_j}{\widetilde{\Psi_i^{\text{std}}}}$, by $\widetilde{\psi}_{i \to j}^{\text{std}}$. It corresponds of course to the transition amplitude from the initial state \(\ket{\Phi_i}\), in which the system is prepared, to the state \(\ket{\Phi_j}\), in which the system may be found at time \(t = 0\), but only up to an overall normalization factor of the dressed state.
These wave functions are represented by sums of diagrams in which intermediate states \(\ket{\Phi_i}\) are excluded.

The state \(\ket{\Psi_i^{\text{std}\,\epsilon}}\) is not, \textit{a priori}, expected to have a definite normalization, as it does not result from the unitary evolution of the normalized state \(\ket{\Phi_i}\). Instead, it results from the action of a given average of unitary operators on $\ket{\Phi_i}$; see Eq.~(\ref{eq:Gell-Mann-averaging}). The limiting state \(\ket{\widetilde{\Psi_i^{\text{std}}}}\) must therefore be normalized a posteriori to restore unitarity. Introducing the probability of observing the system in the state \(\ket{\Phi_i}\),
\begin{equation}
Z_i \equiv \frac{\left| \braket{\Phi_i}{\widetilde{\Psi_i^{\text{std}}}} \right|^2}{\braket{\widetilde{\Psi_i^{\text{std}}}}{\widetilde{\Psi_i^{\text{std}}}}},
\end{equation}
we can rewrite the Fock state expansion~(\ref{eq:standard-PT}), this time for a normalized dressed state, as
\begin{equation}
\frac{\ket{\widetilde{\Psi_i^{\text{std}}}}}{\sqrt{\braket{\widetilde{\Psi_i^{\text{std}}}}{\widetilde{\Psi_i^{\text{std}}}}}} 
= \sqrt{Z_i} \left( \ket{\Phi_i} + \sum_{j \neq i} \widetilde{\psi}_{i \to j}^{\text{std}} \ket{\Phi_j} \right).
\label{eq:Fock-expansion-standard-PT}
\end{equation}
This expression was used, in particular, in the early literature on light-cone perturbation theory; see, e.g., Ref.~\cite{Bjorken:1970ah}. Therein, $Z_i$ is called the wave function renormalization constant. In practice, within this approach, one typically first computes the unnormalized wave functions \(\widetilde{\psi}_{i \to j}^{\text{std}}\) using the perturbation theory formula~(\ref{eq:standard-PT}), and subsequently determines \(Z_i\) by invoking unitarity:
\begin{equation}
Z_i = \frac{1}{1 + \sum_{j \neq i} \left| \widetilde{\psi}_{i \to j}^{\text{std}} \right|^2}.
\label{eq:Zi-from-unitarity}
\end{equation}


\subsubsection{Alternative calculation}

We have presented standard perturbation theory, starting from the physical setup proposed in Ref.~\cite{Gell-Mann:1953dcn}, and formulated it in a way that deliberately parallels the adiabatic switching prescription, to be discussed in the following section. Before proceeding, let us revisit Eq.~(\ref{eq:Gell-Mann-averaging}) and, for completeness, outline an alternative route to the perturbative series. Substituting $U_I$ with Eq.~(\ref{eq:interaction-picture-schrodinger-0}), with ${\cal H}_{0R}\to{\cal H}_0$, setting $t=0$ and integrating over $t_0$, we obtain the formal operator equation\footnote{See also Ref.~\cite{Gell-Mann:1953dcn}, Eq.~(3.29).}
\be
\ket{\Psi_i^{\text{std}\,\epsilon}} = \frac{i\epsilon}{E_i - H + i\epsilon}\ket{\Phi_i}.
\label{eq:time-independent}
\ee
Here, the operator acting on the free eigenstate $\ket{\Phi_i}$ is nothing but the resolvent of the Schrödinger equation, defined with the incoming-wave boundary condition.

We now show how to recover Eq.~(\ref{eq:psi_i_eps-standard}) within this framework. One substitutes $H = \mathcal{H}_{0R} + \mathcal{H}_{1R}$ and expands the operator acting on $\ket{\Phi_i}$ in the following formal series:
\begin{multline}
\frac{i\epsilon}{E_i-H+i\epsilon}=\frac{i\epsilon}{E_i-{\cal H}_{0R}+i\epsilon}+\frac{1}{E_i-{\cal H}_{0R}+i\epsilon}{\cal H}_{1R}\frac{i\epsilon}{E_i-{\cal H}_{0R}+i\epsilon}\\
+\frac{1}{E_i-{\cal H}_{0R}+i\epsilon}{\cal H}_{1R}\frac{1}{E_i-{\cal H}_{0R}+i\epsilon}{\cal H}_{1R}\frac{i\epsilon}{E_i-{\cal H}_{0R}+i\epsilon}+\cdots
\end{multline}
Applying this operator series to the state $\ket{\Phi_i}$, and using the fact that $\ket{\Phi_i}$ is an eigenstate of $\mathcal{H}_{0R}$ with, generically, eigenvalue $\varepsilon_{iR}$, the first term on the right-hand side reproduces $\ket{\Phi_i}$ up to a numerical factor $i\epsilon/(E_i-\varepsilon_{iR}+i\epsilon)$, which is 1 when $\varepsilon_{iR}$ is set to $E_i$. The generic term in the series reads
\begin{multline}
\left(\frac{1}{E_i-{\cal H}_{0R}+i\epsilon}{\cal H}_{1R}\right)^n\frac{i\epsilon}{E_i-{\cal H}_{0R}+i\epsilon}\ket{\Phi_i}\\
=\frac{i\epsilon}{E_i-\varepsilon_{iR}+i\epsilon}\left(\sum_{j,j_{n-1},\cdots,j_1}\ket{\Phi_j}\frac{\mel{\Phi_j}{{\mathcal H}_{1R}}{\Phi_{j_{n-1}}}}{E_i-\varepsilon_{iR}+i\epsilon}\prod_{k=1}^{n-1}\frac{\mel{\Phi_{j_{n-k}}}{{\mathcal H}_{1R}}{\Phi_{j_{n-k-1}}}}{E_i-\varepsilon_{j_{n-k}R}+i\epsilon}\right)
\end{multline}
for $n\geq 1$, and with $j_0\equiv i$.
Summing over $n$, we see that the obtained series,
\begin{multline}
{\ket{\Psi_i^{\text{std}\,\epsilon}}}=\frac{i\epsilon}{E_i-\varepsilon_{iR}+i\epsilon}\Bigg(\ket{\Phi_i}+\sum_{j}\ket{\Phi_j}\frac{\mel{\Phi_j}{{\mathcal H}_{1R}}{\Phi_i}}{E_i-\varepsilon_{jR}+i\epsilon}+\sum_{j,j_1}\ket{\Phi_{j}}\frac{\mel{\Phi_{j}}{{\mathcal H}_{1R}}{\Phi_{j_1}}\mel{\Phi_{j_1}}{{\mathcal H}_{1R}}{\Phi_i}}{(E_i-\varepsilon_{jR}+i\epsilon)(E_i-\varepsilon_{j_1R}+i\epsilon)}\\
+\sum_{j,j_2,j_1}\ket{\Phi_{j}}\frac{\mel{\Phi_{j}}{{\mathcal H}_{1R}}{\Phi_{j_2}}\mel{\Phi_{j_2}}{{\mathcal H}_{1R}}{\Phi_{j_1}}\mel{\Phi_{j_1}}{{\mathcal H}_{1R}}{\Phi_i}}{(E_i-\varepsilon_{jR}+i\epsilon)(E_i-{\varepsilon}_{j_2R}+i\epsilon)(E_i-\varepsilon_{j_1R}+i\epsilon)}+\cdots\Bigg),
\end{multline}
is very close in its form to Eq.~(\ref{eq:psi_i_eps-standard}). Both equations coincide when the Hamiltonian is partitioned such that the energies $\varepsilon_{kR}$ are chosen to match the exact energies $E_k$.

After factorizing the singular wave function $\braket{\Phi_i}{\Psi^{\text{std}\,\epsilon}_i}$, 
we obtain a more general perturbation theory than that we were naturally led to in the previous framework:
\begin{mdframed}[innertopmargin=-5pt, innerbottommargin=5pt, leftmargin=0pt, rightmargin=0pt,skipabove=2pt,skipbelow=2pt]
\begin{multline}
\ket{\widetilde{\Psi_i^{\text{std}}}} =
\ket{\Phi_i}
+ \sum_{j \neq i} \ket{\Phi_j}\Bigg( \frac{1}{E_i - \varepsilon_{jR}} \mel{\Phi_j}{{\cal{H}}_{1R}}{\Phi_i}
+ \sum_{j_1 \neq i} \frac{1}{E_i - \varepsilon_{jR}} \mel{\Phi_{j}}{{\cal{H}}_{1R}}{\Phi_{j_1}} \frac{1}{E_i - \varepsilon_{j_1R}} \mel{\Phi_{j_1}}{{\cal{H}}_{1R}}{\Phi_i} \\
+ \sum_{j_2,j_1 \neq i} \frac{1}{E_i - \varepsilon_{jR}} \mel{\Phi_{j}}{{\cal{H}}_{1R}}{\Phi_{j_2}} \frac{1}{E_i - \varepsilon_{j_2R}} \mel{\Phi_{j_2}}{{\cal{H}}_{1R}}{\Phi_{j_1}} \frac{1}{E_i - \varepsilon_{j_1R}} \mel{\Phi_{j_1}}{{\cal{H}}_{1R}}{\Phi_i}\Bigg)
+ \cdots
\label{eq:standard-PT-bare}
\end{multline}
\end{mdframed}
The energy denominators involve the difference between the physical energy of the asymptotic state and the bare energy of the Fock state onto which one is projecting. This is the perturbation theory formula given in reference papers such as Ref.~\cite{Brodsky:1997de} (see Sec.~3.3 therein). 

When the energies $\varepsilon_{kR}$  are set to $E_k$, we obviously get back Eq.~(\ref{eq:standard-PT}).


\subsection{Adiabatic switching}

We now set aside the standard construction and restart the development of perturbation theory from scratch, this time adopting the ``adiabatic switching prescription.'' In essence, this consists of modifying the theory so that the interaction is slowly, adiabatically, turned off at asymptotic times, leaving the particles effectively free of any interaction in the distant past and future.

\subsubsection{Formulation}

To implement this prescription, we introduce a time dependence in the interaction Hamiltonian \({\cal H}_{1R}\), along with a parameter \(\epsilon \geq 0\) that controls the rate of switching on and off of the interaction. Specifically, we replace the full Hamiltonian \(H\) with a time-dependent one:
\begin{equation}
H^\epsilon(t) \equiv {\mathcal H}_{0R} + {\mathcal H}_{1R}^\epsilon(t),
\end{equation}
with the requirements that $H^{\epsilon=0}(t)=H^\epsilon(t=0)=H$ and \(\lim_{|t| \to \infty} H^{\epsilon>0}(t) = {\mathcal H}_{0R}\). 

There are various ways to achieve this behavior. One may, for instance:
\begin{enumerate}
\renewcommand{\labelenumi}{(\roman{enumi})}
    \item define \({\mathcal H}_{1R}^\epsilon(t) \equiv e^{-\epsilon |t|} {\mathcal H}_{1R}\), or
    \item expand \({\mathcal H}_{1R}\) as \({\mathcal H}_{1R} = \lambda V_1 + \lambda^2 V_2 + \cdots\), where the \(V_j\) are independent of \(\lambda\), and replace the coupling constant \(\lambda\) with the time-dependent factor \(\lambda e^{-\epsilon |t|}\), yielding
    \begin{equation}
    {\mathcal H}_{1R}^\epsilon(t) \equiv \lambda\, e^{-\epsilon |t|} V_1 + \lambda^2 e^{-2\epsilon |t|} V_2 + \cdots
    \end{equation}
\end{enumerate}
These two prescriptions coincide when \({\mathcal H}_{1R}\) depends linearly on \(\lambda\), but otherwise they differ --- for instance, in non-Abelian gauge theories or whenever $\lambda$-dependent counterterms must be introduced.
In both prescriptions, at sufficiently large times \(|t|\gg 1/\epsilon\), the system is effectively governed by the free Hamiltonian \({\mathcal H}_{0R}\), with energy levels given by its eigenvalues \(\varepsilon_{iR}\). As the interaction is slowly, or ``adiabatically,'' turned on while the system approaches the probing time $t=0$, these energies evolve toward the physical ones, namely the eigenvalues \(E_i\) of the full Hamiltonian \(H\).

Let us now examine the perturbation theory that follows from these adiabatic prescriptions.

\subsubsection{Perturbative series}

The Hamiltonian in the Schrödinger equation~(\ref{eq:basic-schrodinger}) is now time-dependent. Let us write the latter directly for the Schrödinger-picture evolution operator. For $t_0\leq t$,
\be
i\frac{\partial}{\partial t} U_S^{\text{ad}\,\epsilon}(t,t_0)=H^\epsilon(t)U_S^{\text{ad}\,\epsilon}(t,t_0),
\quad \text{with}\quad U_S^{\text{ad}\,\epsilon}(t=t_0,t_0)={\mathbbm I}.
\label{eq:schrodinger-ad-forward}
\ee
The solution to this equations formally writes
\be
U_S^{\text{ad}\,\epsilon}(t,t_0)\equiv T\exp\left(-i\int_{t_0}^{t}dt' {H}^\epsilon(t')\right).
\ee
It is straightforward to check that it also satisfies the backward evolution
\be
-i\frac{\partial}{\partial{t_0}} U_S^{\text{ad}\,\epsilon}(t,t_0)=U_S^{\text{ad}\,\epsilon}(t,t_0)H^\epsilon(t_0),
\quad \text{with}\quad U_S^{\text{ad}\,\epsilon}(t,t_0=t)={\mathbbm I},
\label{eq:schrodinger-ad-backward}
\ee
that we will need later on.

To express the evolution operator in the interaction picture, one proceeds as in the standard perturbation theory. Transforming the Schrödinger-picture evolution operator just written down, one finds
\be
U_{IR}^{\text{ad}\,\epsilon}(t,t_0)\equiv e^{it{\mathcal H}_{0R}} U_S^{\text{ad}\,\epsilon}(t,t_0) e^{-it_0{\mathcal H}_{0R}}.
\ee
Alternatively, solving the interaction-picture Schrödinger equation~(\ref{eq:interaction-picture-schrodinger}), one gets 
\be
U_{IR}^{\text{ad}\,\epsilon}(t,t_0)=T\exp\left(-i\int_{t_0}^{t}dt' {\mathcal H}_{IR}^\epsilon(t')\right),\quad\text{with}\quad
{\mathcal H}_{IR}^\epsilon(t')=e^{it'{\mathcal H}_{0R}}{\mathcal H}_{1R}^\epsilon(t') e^{-it'{\mathcal H}_{0R}}.
\ee
For any fixed $\epsilon$, the operator $U_{IR}^{\text{ad}\,\epsilon}(t,t_0)$ in which $t$ or $t_0$ are sent to positive or negative infinity exists.

Let us start with prescription (i). Specializing to $t=0$ and $t_0=-\infty$, the series expansion of the evolution operator reads
\be
U_{IR}^{\text{ad}\,\epsilon}(0,-\infty)={\mathbbm I}-i\int_{-\infty}^0 dt\,{\mathcal H}_{IR}^\epsilon(t)+(-i)^2\int_{-\infty}^0 dt_2\,{\mathcal H}_{IR}^\epsilon(t_2)\int_{-\infty}^{t_2} dt_1\,{\mathcal H}_{IR}^\epsilon(t_1)+\cdots
\ee
We apply this operator to $\ket{\Phi_i}$ to get the dressed state:
\be
\ket{\Psi^{\text{ad}\,\epsilon}_i}\equiv
U_{IR}^{\text{ad}\,\epsilon}(0,-\infty)\ket{\Phi_i}.
\label{eq:def-Psi-ad}
\ee
Using repeatedly the completeness relation~(\ref{eq:completeness}) and performing the time integrations, we find that the Fock-state expansion of the dressed state reads
\begin{multline}
\ket{\Psi^{\text{ad}\,\epsilon}_i}=\sum_j\ket{\Phi_j}\Bigg(\delta_{ji}+\frac{1}{\varepsilon_{iR}-\varepsilon_{jR}+i\epsilon}\mel{\Phi_j}{{\mathcal H}_{1R}}{\Phi_i}+\sum_{j_1}\frac{1}{\varepsilon_{iR}-\varepsilon_{jR}+2i\epsilon}\frac{\mel{\Phi_{j}}{{\mathcal H}_{1R}}{\Phi_{j_1}}\mel{\Phi_{j_1}}{{\mathcal H}_{1R}}{\Phi_i}}{\varepsilon_{iR}-\varepsilon_{j_1R}+i\epsilon}\\
+\sum_{j_2,j_1}\frac{1}{\varepsilon_{iR}-\varepsilon_{jR}+3i\epsilon}\frac{\mel{\Phi_{j}}{{\mathcal H}_{1R}}{\Phi_{j_2}}\mel{\Phi_{j_2}}{{\mathcal H}_{1R}}{\Phi_{j_1}}\mel{\Phi_{j_1}}{{\mathcal H}_{1R}}{\Phi_i}}{(\varepsilon_{iR}-\varepsilon_{j_2R}+2i\epsilon)(\varepsilon_{iR}-\varepsilon_{j_1R}+i\epsilon)}+\cdots\Bigg).
\label{eq:Fock-expansion-i}
\end{multline}
At variance with Eq.~(\ref{eq:psi_i_eps-standard}), the $i\epsilon$'s in the denominators now come with different integer factors. Moreover, the energies come out naturally as the eigenvalues of the renormalized free Hamiltonian ${\mathcal H}_{0R}$, not of the full Hamiltonian~$H$.

As for prescription (ii), the formula is slightly more complicated. It is convenient to order the perturbative series according to the number of energy denominators:
\begin{mdframed}[innertopmargin=-5pt, innerbottommargin=5pt, leftmargin=0pt, rightmargin=0pt,skipabove=2pt,skipbelow=2pt]
\begin{multline}
\ket{\Psi^{\text{ad}\,\epsilon}_i}=\sum_j\ket{\Phi_j}\Bigg(\delta_{ji}+\sum_{n\geq1}\frac{1}{\varepsilon_{iR}-\varepsilon_{jR}+ni\epsilon}\mel{\Phi_j}{\lambda^n V_n}{\Phi_i}\\
+\sum_{j_1}\sum_{n,n_1\geq1}\frac{1}{\varepsilon_{iR}-\varepsilon_{jR}+(n+n_1)i\epsilon}\frac{\mel{\Phi_{j}}{\lambda^n V_{n}}{\Phi_{j_1}}\mel{\Phi_{j_1}}{\lambda^{n_1}V_{n_1}}{\Phi_i}}{\varepsilon_{iR}-\varepsilon_{j_1R}+n_1i\epsilon}+\cdots\Bigg).
\label{eq:Fock-expansion-ii}
\end{multline}
\end{mdframed}
We shall denote by $\psi^{\text{ad}\,\epsilon}_{i\to j}$ the coefficient of the Fock state $\ket{\Phi_j}$ in this series.

Most of the literature that uses adiabatic switching assumes an interaction Hamiltonian linear in the coupling. In such a case, turning off the interaction Hamiltonian or the coupling at asymptotic times, namely using prescriptions (i) or (ii), is identical. In Ref.~\cite{Mueller:2012bn} in which a next-to-leading-order calculation was performed using adiabatic switching, prescription (ii) was tacitly assumed. As our goal is to examine how perturbative calculations are organized in the adiabatic switching prescription as compared to the standard one, we will from now make a choice, and adhere to prescription (ii). The latter is technically more natural, essentially because powers of $\lambda$ naturally organize the expansion, so damping $\lambda$ aligns the adiabatic regulator with the perturbative expansion. This will become especially evident in the next section, in the proof of the Gell-Mann--Low theorem beyond the linear case.

At variance with $\ket{\Psi_i^\epsilon}$ defined in Eq.~(\ref{eq:Gell-Mann-averaging}), the state $\ket{\Psi_i^{\text{ad}\,\epsilon}}$ is normalized like $\ket{\Phi_i}$ by construction, as is clear from its definition in Eq.~(\ref{eq:def-Psi-ad}). However, in general, the limit $\epsilon\to 0$ is still divergent. Indeed, one observes that energy denominators may tend to zero, notably when intermediate states coincide with the asymptotic state. As we shall see, these singularities resum into a global phase factor, with a phase that diverges as $1/\epsilon$ when $\epsilon \to 0$. Once this factor is divided out, the resulting state is an eigenstate of the full Hamiltonian, as we will now show.


\subsubsection{The Gell-Mann and Low theorem}

Similarly to $\ket{\widetilde{\Psi_i^{\text{std}\,\epsilon}}}$ introduced in the context of the standard prescription, we define
\be
\ket{\widetilde{\Psi_i^{\text{ad}\,\epsilon}}}\equiv \frac{\ket{\Psi_i^{\text{ad}\,\epsilon}}}{\braket{\Phi_i}{\Psi_i^{\text{ad}\,\epsilon}}}.
\label{eq:def-finite-eigen}
\ee
The Gell-Mann and Low theorem \cite{Gell-Mann:1951ooy} states that if the $\epsilon\to 0$ limit of the one-parameter family of vectors $\left\{\ket{\widetilde{\Psi_i^{\text{ad}\,\epsilon}}}\right\}_{\epsilon}$ exists, then the limit state
\be
\ket{\widetilde{\Psi_i^{\text{ad}}}}\equiv\lim_{\epsilon\to 0}\ket{\widetilde{\Psi_i^{\text{ad}\,\epsilon}}}
\ee
is an eigenstate of the Hamiltonian $H$.

The proof of this theorem was originally given in the Appendix of Ref.~\cite{Gell-Mann:1951ooy} (see also textbooks such as Ref.~\cite{Fetter:2003}), formulating perturbation theory using adiabatic switching. The key of the proof lies in the following identity:
\be
\left(H-\varepsilon_{iR}-i\epsilon \lambda\frac{\partial}{\partial \lambda}\right)\ket{\Psi_i^{\text{ad}\,\epsilon}}=0\,.
\label{eq:G-M&L}
\ee
The latter was originally established using a power expansion of the evolution operator. A simpler re-derivation was proposed more recently~\cite{Molinari:2007}. It was assumed that the interaction terms in $H$ depend linearly on the coupling. But it is straightforward to extend the reasoning to a Hamiltonian in which the coupling appears in different powers, provided adiabatic switching is implemented as in prescription (ii). Let us slightly reformulate the proof of Ref.~\cite{Molinari:2007}, incorporating this extension.

Defining $\theta\equiv (\ln\lambda)/\epsilon$, we write the Schrödinger-picture evolution operator within prescription (ii) as
\be
U_S^{\text{ad}\,\epsilon}(t,t_0)\equiv T\exp\left[-i\int_{t_0}^{t}dt' \bigg({\cal H}_{0R}+\sum_{k\geq 1} e^{k\epsilon (\theta+t')}V_k\bigg)\right]=T\exp\left[-i\int_{t_0+\theta}^{t+\theta}dt' \bigg({\cal H}_{0R}+\sum_{k\geq 1} e^{k\epsilon t'}V_k\bigg)\right]\,.
\ee
From the first to the second expression, we have merely performed a translation of the integration variable $t'$. The resulting expression for the evolution operator shows that its derivative with respect to $\theta$ corresponds to an appropriate combination of time derivatives:
\be
\frac{\partial}{\partial\theta}U_S^{\text{ad}\,\epsilon}(t,t_0)=\frac{\partial}{\partial t}U_S^{\text{ad}\,\epsilon}(t,t_0)+\frac{\partial}{\partial t_0}U_S^{\text{ad}\,\epsilon}(t,t_0)\,.
\ee
Replacing $\partial_\theta$ by $\epsilon\lambda\partial_\lambda$ in the left-hand side and using the appropriate Schrödinger equation [see Eqs.~(\ref{eq:schrodinger-ad-forward}),(\ref{eq:schrodinger-ad-backward})] to express the two terms in the right-hand side, we get
\be
i\epsilon\lambda\frac{\partial}{\partial\lambda}U_S^{\text{ad}\,\epsilon}(t,t_0)=H^\epsilon(t)U_S^{\text{ad}\,\epsilon}(t,t_0)-U_S^{\text{ad}\,\epsilon}(t,t_0)H^\epsilon(t_0)\,.
\ee
We convert this equation to the interaction picture, by multiplying it left by $e^{it{\cal H}_{0R}}$, and right by $e^{-it_0{\cal H}_{0R}}$. Setting $t=0$ and letting $t_0$ go to $-\infty$:\footnote{As a side remark, we can observe that if the left-hand side vanishes, this equation says that $U_{IR}^{\text{ad}\,\epsilon}(0,-\infty)$ is an intertwiner of the free and the full Hamiltonians. This is a basic property of M{\o}ller operators.}
\be
i\epsilon\lambda\frac{\partial}{\partial\lambda}U_{IR}^{\text{ad}\,\epsilon}(0,-\infty)=HU_{IR}^{\text{ad}\,\epsilon}(0,-\infty)-U_{IR}^{\text{ad}\,\epsilon}(0,-\infty){\cal H}_{0R}.
\ee
Lastly, multiplying right by $\ket{\Phi_i}$ and rearranging the terms, we arrive at Eq.~(\ref{eq:G-M&L}).

The next step is turning Eq.~(\ref{eq:G-M&L}) into an equation for $\ket{\widetilde{\Psi_i^{\text{ad}\,\epsilon}}}$. To this aim, we first divide Eq.~(\ref{eq:G-M&L}) by $\braket{\Phi_i}{\Psi_i^{\text{ad}\,\epsilon}}$:
\be
\left(H-\varepsilon_{iR}\right)\ket{\widetilde{\Psi_i^{\text{ad}\,\epsilon}}}-i\epsilon \frac{1}{\braket{\Phi_i}{\Psi_i^{\text{ad}\,\epsilon}}}\lambda\frac{\partial}{\partial \lambda}\ket{\Psi_i^{\text{ad}\,\epsilon}}=0\,,
\label{eq:G-M&L-norm}
\ee
where we have used the definition~(\ref{eq:def-finite-eigen}). We then contract this equation left with $\bra{\Phi_i}$. Introducing
\be
E_i^{\epsilon}\equiv\mel{\Phi_i}{H}{\widetilde{\Psi_i^{\text{ad}\,\epsilon}}}\,,
\ee
we get the following equation:
\be
E_i^{\epsilon}-\varepsilon_{iR}-i\epsilon\lambda\frac{\partial}{\partial\lambda}\ln\braket{\Phi_i}{\Psi_i^{\text{ad}\,\epsilon}}=0\,.
\label{eq:rel-E-epsilon}
\ee
Multiplying the latter by $\ket{\widetilde{\Psi_i^{\text{ad}\,\epsilon}}}$ and subtracting the resulting equation from Eq.~(\ref{eq:G-M&L-norm}), we arrive at
\be
\left(H-E^{\epsilon}_{i}-i\epsilon \lambda\frac{\partial}{\partial \lambda}\right)\ket{\widetilde{\Psi_i^{\text{ad}\,\epsilon}}}=0\,.
\label{eq:eq-phase}
\ee

Let us now assume that the limit state $\ket{\widetilde{\Psi_i^{\text{ad}}}}$ indeed exists. Then $E^{\epsilon}_{i}$ tends to a finite value $E_i$ when $\epsilon\to 0$. In this limit, the previous equation boils down to the stationary Schrödinger equation $H\ket{\widetilde{\Psi_i^{\text{ad}}}}=E_i\ket{\widetilde{\Psi_i^{\text{ad}}}}$, showing that $\ket{\widetilde{\Psi_i^{\text{ad}}}}$ is an eigenvector of $H$ with eigenvalue $E_i$. Since it is, by construction, the dressed state of $\ket{\Phi_i}$, we can write that all the following state vectors are proportional to each other:
\be
\ket{\widetilde{\Psi_i^{\text{ad}}}}\propto\ket{\widetilde{\Psi_i^{\text{std}}}}\propto\ket{\Psi_i}.
\ee

\paragraph{Energy shift from the phase of the dressed state}

The inner product ${\braket{\Phi_i}{\Psi_i^{\text{ad}\,\epsilon}}}$, namely $\psi_{i\to i}^{\text{ad}\,\epsilon}$, is the probability amplitude of finding the dressed state $\ket{{\Psi_i^{\text{ad}\,\epsilon}}}$ in the corresponding bare state $\ket{\Phi_i}$. It is thus a wave function of the system, namely a complex function of $\epsilon$ and $\lambda$, and possibly other parameters. Let us write it as the product of a modulus and phase factor
\be
\psi_{i\to i}^{\text{ad}\,\epsilon}=e^{{\alpha_i(\epsilon,\lambda)}/{(i\epsilon)}}r_i(\epsilon,\lambda),
\ee
where $\alpha_i$ and $r_i$ are real, the latter being chosen positive. In terms of these quantities, Eq.~(\ref{eq:rel-E-epsilon}) writes
\be
E_{i}^\epsilon-\varepsilon_{iR}=\lambda\frac{\partial}{\partial\lambda}\alpha_i(\epsilon,\lambda)+i\epsilon\lambda\frac{\partial}{\partial\lambda}\ln\,r_i(\epsilon,\lambda)\,.
\ee
The left-hand side tends to the real number $E_i-\varepsilon_{iR}$ when $\epsilon\to0$. Necessarily, the first term in the right-hand side must be regular in this limit, and the second term needs to vanish. This equation then boils down to
\begin{equation}
E_i-\varepsilon_{iR}=\lim_{\epsilon\to 0}\lambda\frac{\partial}{\partial\lambda}\alpha_i(\epsilon,\lambda).
\label{eq:energy-shift-GML}
\end{equation}
This formula means that the energy shift between the bare and physical energies, that is, the difference between the eigenvalues of $H_0$ and the corresponding eigenvalues of $H$, is given by the $\epsilon \to 0$ limit of the derivative of the phase of the wave function with respect to the logarithm of the coupling. Note that it implies that $\alpha_i$ must be independent of $\lambda$ when the renormalized energy $\varepsilon_{iR}$ is chosen to match the physical energy $E_i$. Clearly, $\alpha_i(\epsilon, \lambda = 0) = 0$, since the phase is generated by the interaction. It follows that $\alpha_i$ vanishes identically if $\varepsilon_{iR}\equiv E_i$.

One may actually see quite easily that, up to a sign, the same phase factor is obtained for any wave function ${\braket{\Phi_j}{\Psi_i^{\text{ad}\,\epsilon}}}$. Hence the Fock-state expansion of the dressed state $\ket{\Psi_i^{\text{ad}\,\epsilon}}$ normalized by any of the latter, or out of which the phase factor has been divided, 
consists of a real superposition of the bare Fock state vectors $\ket{\Phi_k}$.

\subsubsection{Fock state expansion}

The form of the Fock state expansion of the eigenstates of \(H\) in this formulation of perturbation theory is \textit{a priori} quite different from the standard one~(\ref{eq:Fock-expansion-standard-PT}):
\begin{equation}
\ket{\check{\Psi}_i^{\text{ad}}}\equiv
\lim_{\epsilon \to 0} \left( e^{-\alpha_i(\epsilon, \lambda) / (i \epsilon)} \ket{\Psi_i^{\text{ad}\, \epsilon}} \right)
=\sum_j \check{\psi}^\text{ad}_{i \to j} \ket{\Phi_j}\,.
\label{eq:Fock-expansion-adiabatic}
\end{equation}
It is worth noting that such a factorization of singularities as $\epsilon \to 0$ was already discussed in Refs.~\cite{Toth:2008un,Toth:2009gh}, in the context of defining a singularity-free $S$-matrix.

The key difference between~(\ref{eq:Fock-expansion-standard-PT}) and (\ref{eq:Fock-expansion-adiabatic}) is that the wave functions \(\check{\psi}^\text{ad}_{i \to j}\) are already normalized in this formulation, meaning that their moduli squared are proper probabilities. The correspondence with the wave functions computed in standard perturbation theory is as follows:
\begin{equation}
\check{\psi}^\text{ad}_{i \to i} = \sqrt{Z_i}
\quad \text{and} \quad
\frac{\check{\psi}^\text{ad}_{i \to j}}{\check{\psi}^\text{ad}_{i \to i}} = \widetilde{\psi}_{i \to j}^{\text{std}}.
\label{eq:correspondence}
\end{equation}


\section{State vector of a particle in models with no space dimension}
\label{sec:0D}

In this section, we illustrate the properties of perturbative expansions within the adiabatic switching prescription using the simplest quantum mechanical model we can think of, one that allows explicit computations to arbitrary orders.

\subsection{Definition of the model}

\subsubsection{Hamiltonian and Hilbert space}

We consider the following general Hamiltonian:
\be
H=\sum_{i,j\geq 1}\lambda_{ij}(a^\dagger)^i a^j
\ee
where $a$ is the annihilation operator and $a^\dagger$ the creation operator, Hermitian conjugate to each other, and obeying the canonical commutation relation $[a,a^\dagger]=1$, $[a,a]=[a^\dagger,a^\dagger]=0$. Starting from the vacum $\ket{0}$ which is such that $a\ket{0}=0$, we build $n$-particle states by acting $n$ times $a^\dagger$ on the vacuum, with the following normalization:
\be
\ket{n}=\frac{\left(a^\dagger\right)^n}{\sqrt{n!}}\ket{0}.
\ee
The action of $a$ and $a^\dagger$ on $\ket{n}$ obviously read
\be
a\ket{n}=\sqrt{n}\ket{n-1},\quad a^\dagger\ket{n}=\sqrt{n+1}\ket{n+1}.
\ee
The states satisfy the completeness and orthogonality relations in the form
\be
\sum_n\ket{n}\bra{n}={\mathbbm I},\quad \braket{m}{n}=\delta_{m,n}.
\ee

Here, we set all $\lambda_{ij}$'s to zero, except $\lambda_{11}\equiv\omega$ and $\lambda_{12}=\lambda_{21}\equiv\lambda$, where $\omega$ and $\lambda$ are real, $\omega>0$. Then, the Hamiltonian reads\footnote{Note that our Hamiltonian corresponds to that of Reggeon Field Theory in zero spatial dimensions, provided that $1+\omega$ is identified with the intercept of the Regge trajectory, and that $\lambda_{12} = \lambda_{21}$ is set to $i g/2$, with $g$ a real coupling constant. In this case, however, the Hamiltonian is not Hermitian, and time evolution is therefore not unitary. Such models have been investigated in the literature, particularly in the context of high-energy scattering; see, e.g., Refs.~\cite{Shoshi:2005pf,Bondarenko:2006rh}. Incidentally, one of the resolution methods in~\cite{Shoshi:2005pf}, called ``the $\omega$-representation,'' amounts to perturbation theory in the standard prescription.}
\be
H=\underbrace{\omega \, a^\dagger a}_{\equiv H_0}+\underbrace{\lambda\left(a^{\dagger}a^\dagger a+a^\dagger a a\right)}_{\equiv H_1}
\ee
Although not strictly necessary in the context of this quantum mechanical model -- as opposed to the case of a generic field theory-- we may nevertheless choose to parametrize the Hamiltonian using an ad hoc frequency $\omega_R$, rather than the original, ``bare'' frequency $\omega$. We then introduce the notation $Z_m$ for their ratio, $\omega / \omega_R \equiv Z_m$. This quantity depends \textit{a priori} on the two parameters $\lambda$ and $\omega_R$. For $H$ to remain the same after this reparametrization, we need to add a compensating term $(Z_m-1)\omega_R a^\dagger a$. We then split $H$ into a free and interaction part as $H\equiv {\mathcal H}_{0R}+{\mathcal H}_{1R}$, with
\be
\begin{aligned}
{\mathcal H}_{0R}&={\omega_R\, a^\dagger a}\\
{\mathcal H}_{1R}&={\lambda\left(a^{\dagger}a^\dagger a+a^\dagger a a\right)+(Z_m-1)\omega_R\, a^\dagger a}\,.
\end{aligned}
\ee
Expanding the renormalization constant $Z_m$ in power series of the coupling:
\be
\begin{aligned}
&Z_m=1+\sum_{k=1}^\infty \lambda^{2k}Z_m^{(2k)}(\omega_R) \implies
{\mathcal H}_{1R}=\lambda V_1+\sum_{k=1}^\infty \lambda^{2k} V_{2k},\\
&\text{with}\quad
V_1\equiv {a^{\dagger}a^\dagger a+a^\dagger a a}
\quad\text{and}\quad 
V_{2k}=Z_m^{(2k)}(\omega_R)\,\omega_R\, a^\dagger a\,.
\end{aligned}
\label{eq:Ansatz-H-0D}
\ee
We have anticipated that we can restrict ourselves to $Z_m$'s defined as a series in even powers of $\lambda$ only. In practice, the values we shall choose for $\omega_R$ will be either the bare energy $\omega$, or the energy $\omega_P$ of the (perturbative) ground state of the full Hamiltonian $H$. 

As announced, we will use the adiabatic switching prescription (ii), hence promote the coupling $\lambda$ to $\lambda e^{-\epsilon|t|}$. To simplify the notation, we will omit the subscripts and superscripts ``$I$'', ``$R$'', and ``$\epsilon$'' on operators, states, and wave functions.

\subsubsection{Diagrammatic calculation rules for wave functions}

To evaluate the contributions to the wave function of a single initial particle, we use the perturbation theory formula~(\ref{eq:Fock-expansion-ii}). 
The eigenstates $\ket{\Phi_n}$ of the free Hamiltonian ${\mathcal H}_{0R}$ in there obviously correspond to $\ket{n}$ and have respective energies $\varepsilon_{nR}=n\times\omega_R$. 
The rules for the energy denominators is the following. If the $j$-th transition, in time-order starting from $t=-\infty$, brings the system to the state $\ket{\Phi_n}$, then it comes with the energy denominator factor
\be
\frac{1}{\varepsilon_{1R}-\varepsilon_{nR}+il\epsilon}=\frac{1}{(1-n)\omega_R +il\epsilon}\ ,
\label{eq:denom}
\ee
where $l\geq j$ is the total number of powers of $\lambda$ accumulated in the successive transitions numbered $1,2,\cdots,j$.

The potentially non-zero matrix elements of ${\mathcal H}_{1R}$ read
\be
\begin{aligned}
&\mel{n}{{\mathcal H}_{1R}}{n-1}=\mel{n-1}{{\mathcal H}_{1R}}{n}=\lambda(n-1)\sqrt{n}\ ,\\
&\mel{n}{{\mathcal H}_{1R}}{n}=n\,\omega_R\left(Z_m-1\right)= n\,\omega_R\sum_{k=1}^\infty\lambda^{2k}Z_m^{(2k)}(\omega_R)\,.
\end{aligned}
\label{eq:H1prime}
\ee


\subsection{Fock state expansion in the adiabatic switching prescription: restricted Fock space}

We shall examine in details the wave functions obtained when the Fock space is restricted to the vector states $\ket{1}$ and $\ket{2}$.

\subsubsection{One loop}
Let us begin by computing transition probabilities to order \(\lambda^2\). At this order, we must include contributions from both one- and two-particle Fock states.

The two nontrivial contributions to the \(1 \to 1\) transition amplitude are given below, each accompanied by a diagrammatic illustration:
\begin{equation}
\leg\oneloop\leg = \frac{\lambda \sqrt{2}}{-\omega_R + i \epsilon} \times \frac{\lambda \sqrt{2}}{2 i \epsilon}
\quad\quad\quad
\leg\ctmass{2}\leg = \frac{Z_m^{(2)} \omega_R \lambda^2}{2 i \epsilon}
\end{equation}
The number 2 above the second diagram representing the counterterm vertex contribution indicates that the vertex is of order \(\lambda^2\).
Keeping only the terms that survive in the limit \(\epsilon \to 0\), we find:
\begin{equation}
\psi_{1 \to 1}^\text{ad} = 1 - \frac{\lambda^2}{\omega_R^2} \left[ \left( 1 - \frac{Z_m^{(2)} \omega_R^2}{2} \right) \frac{\omega_R}{i \epsilon} + 1 \right] + o(\lambda^2)
\end{equation}

The wave function of the 2-particle Fock state is given by a unique diagram at the considered order~$\lambda$:
\be
\psi_{1\to2}^{\text{ad}}=\leg\halfcircle =\frac{\lambda\sqrt{2}}{-\omega_R +i\epsilon}+o(\lambda^2)\xrightarrow[\epsilon\to0]{}-\frac{\lambda\sqrt{2}}{\omega_R}+o(\lambda^2).
\label{eq:1to2-O1}
\ee
Unitarity is easily verified at this order:
\be
\left|\psi_{1\to1}^\text{ad}\right|^2+\left|\psi_{1\to2}^\text{ad}\right|^2=1+{o(\lambda^2)}.
\ee

\subsubsection{One loop resummed}

It is instructive to iterate the one-loop diagrams and sum them to all orders in the coupling $\lambda$, in the spirit of leading-logarithmic resummations in gauge theories. This procedure allows us to illustrate several key features of amplitude structures within the adiabatic switching prescription, while keeping the computations relatively elementary.

In doing so, we effectively restrict the Fock space to the states $\ket{1}$ and $\ket{2}$, corresponding to the two lowest-energy levels, while still including contributions at arbitrarily high orders in perturbation theory. One diagram of order $\lambda^4$ that needs to be accounted for reads
\be
\leg\oneloop\leg\oneloop\leg=\frac{\lambda\sqrt{2}}{-\omega_R +i\epsilon}\times\frac{\lambda\sqrt{2}}{2i\epsilon}\times\frac{\lambda\sqrt{2}}{-\omega_R +3i\epsilon}\times\frac{\lambda\sqrt{2}}{4i\epsilon}
\label{eq:one-loop-resum}
\ee
It exhibits $1/\epsilon^2$ and $1/\epsilon$ singularities in the limit $\epsilon\to 0$. 

It is not difficult to compute any iteration of the one-loop diagram. If we resum all of them and include the counterterms, a calculation represented graphically as
\be
\leg\leg\leg+\underbrace{\left(\leg\oneloop\leg+\leg\ctmass{2}\leg\right)}_{O(\lambda^2)}+\underbrace{\left[\left(\leg\oneloop\leg+\leg\ctmass{2}\leg\right)\!\times\!\left(\leg\oneloop\leg+\leg\ctmass{2}\leg\right)\right]}_{O(\lambda^4)}+\cdots
\ee
we get
\be
\psi_{1\to 1}^{\text{ad}\,(2\ \text{resum})}\equiv 1+\sum_{n=1}^\infty \frac{\lambda^{2n}}{(i\epsilon\,\omega_R)^n n!}\prod_{k=0}^{n-1} \left(-\frac{1}{1-(2k+1)i\epsilon/\omega_R}+\frac{Z_m^{(2)}\omega_R^2}{2}\right).
\ee
The most singular term at order $2n$ in $\lambda$ is \textit{a priori} proportional to $\lambda^{2n}/(i\epsilon\omega_R)^n$. The set of these terms resum as
\be
\left.\psi_{1\to 1}^{\text{ad}\,(2\ \text{resum})}\right|_\text{most singular}=\exp\left[{\frac{\lambda^2}{i\epsilon\omega_R}\left(\frac{Z_m^{(2)}\omega_R^2}{2}-1\right)}\right].
\ee
The next-to-most singular terms have at least one more power of $\lambda^2$ for each power of $1/\epsilon$. 

From the phase $\alpha(\lambda,\omega_R)\equiv(\lambda^2/\omega_R)(Z_m^{(2)}\omega_R^2/2-1)$, we may compute the shift between the energy parameter $\omega_R$ and the physical energy $\omega_P$ of the lowest-lying state, as a direct application of Eq.~(\ref{eq:energy-shift-GML}):
\be
\omega_P-\omega_R=\lambda\frac{\partial\alpha(\lambda,\omega_R)}{\partial\lambda}\implies
\omega_P=\omega_R\left(1+Z_m^{(2)}\lambda^2-\frac{2\lambda^2}{\omega_R^2}\right)+o(\lambda^2).
\ee 
Let us note that 
\begin{itemize}
    \item setting $Z_m^{(2)}=0$, namely identifying $\omega_R$ to $\omega$, this equation gives the physical energy $\omega_P$ as a function of the bare energy $\omega$. It is straightforward to check that $\omega_P$ is simply the smallest eigenvalue of the restricted Hamiltonian
    \be
    \left.H\right|_{\ket{1},\ket{2}}=\left(\begin{matrix}\omega & \lambda\sqrt{2}\\\lambda\sqrt{2}&2\omega\end{matrix}\right)
    \ee
    expanded to order $\lambda^2$;
    \item setting $Z_m^{(2)}=2/\omega_R^2$, which amounts to identifying $\omega_R$ to $\omega_P$, the phase $\alpha$ vanishes, and the wave function becomes perfectly regular as $\epsilon\to 0$. The series then resums as
    \be
\left.\psi_{1\to 1}^{\text{ad}\,(2\ \text{resum})}\right|_{Z_m^{(2)}=2/\omega_R^2}=\sum_{n=0}^\infty \left(-\frac{\lambda^{2}}{2\omega_R^2}\right)^n\frac{(2n)!}{(n!)^2}=\frac{1}{\sqrt{1+2\lambda^2/\omega_R^2}}
\ee
which coincides with
\be
\frac{1}{\sqrt{1+\left|\psi_{1\to2}^{\text{ad}\,(1)}\right|^2}};
\ee
compare to $\sqrt{Z_1}$, with $Z_1$ computed from Eq.~(\ref{eq:Zi-from-unitarity}), the sum over Fock states being restricted to $j\leq 2$, and $\widetilde\psi_{1\to 2}^\text{std}$ replaced by its lowest-order contribution, denoted by $\psi_{1\to2}^{\text{ad}\,(1)}$ and consisting of the term of order $\lambda$ in Eq.~(\ref{eq:1to2-O1}). 
\end{itemize}


\subsection{All-order analysis across the complete Fock space}

\subsubsection{Exact higher orders}

Let us list the diagrams contributing to order $\lambda^4$. As for the $1\to 1$ wave function, in addition to the diagram~(\ref{eq:one-loop-resum}) (plus counterterms), we have
\be
\begin{aligned}
4\times\quad&\leg\twoloopsint\leg&&=\frac{\lambda\sqrt{2}}{-\omega_R +i\epsilon}\times\frac{\lambda2\sqrt{3}}{-2\omega_R +2i\epsilon}\times\frac{\lambda2\sqrt{3}}{-\omega_R +3i\epsilon}\times\frac{\lambda\sqrt{2}}{4i\epsilon}
\end{aligned}
\ee
The factor 4 in the left-hand side accounts for the four different ways of hooking the internal line. This factor is already included in the matrix elements of ${\cal H}_1$ used in the right-hand side. The corresponding counterterm diagrams are the following:
\be
\begin{aligned}
2\times&\quad\twoloopsct&&=\frac{\lambda\sqrt{2}}{-\omega_R+i\epsilon}\times\frac{\lambda^2 \,2\omega_R Z_m^{(2)}}{-\omega_R+3i\epsilon}\times\frac{\lambda\sqrt{2}}{4i\epsilon},\\
&\quad\quad\ \leg\ctmass{4}\leg&&=\ \frac{\lambda^4 \omega_R Z_m^{(4)}}{4i\epsilon}
\end{aligned}
\ee
In general, all these contributions to the amplitude are singular when $\epsilon\to 0$.

As for the $1\to 2$ wave function, some diagrams are regular, 
\be
\begin{aligned}
4\times\quad & \leg\halfcircleandloop &&=\frac{\lambda\sqrt{2}}{-\omega_R +i\epsilon}\times\frac{\lambda2\sqrt{3}}{-2\omega_R +2i\epsilon}\times\frac{\lambda2\sqrt{3}}{-\omega_R +3i\epsilon}\\
2\times\quad &\onetotwoct &&=\frac{\lambda\sqrt{2}}{-\omega_R+i\epsilon}\times\frac{\lambda^2 2\omega_RZ_m^{(2)}}{-\omega_R+3i\epsilon}
\label{eq:1to2-O3-regular}
\end{aligned}
\ee
and others are singular, due to the presence of single-particle intermediate states:
\be
\begin{aligned}
&\leg\oneloop\leg\halfcircle  &&=\frac{\lambda\sqrt{2}}{-\omega_R +i\epsilon}\times\frac{\lambda\sqrt{2}}{2i\epsilon}\times\frac{\lambda\sqrt{2}}{-\omega_R+3i\epsilon}\\
&\leg\ctmass{2}\leg\halfcircle &&= \frac{\lambda^2\omega_RZ_m^{(2)}}{2i\epsilon}\times\frac{\lambda\sqrt{2}}{-\omega_R+3i\epsilon}
\end{aligned}
\ee
There is also a set of $1\to 3$ diagrams at this order.


The calculation of wave functions to any order in $\lambda$ is actually easy to automatize. We check explicitly that we may write the Fock-state expansion of the interaction state $\ket{\mathbbm i}$ of a single initial particle in the limit $\epsilon\to 0$ as prescribed in Eq.~(\ref{eq:Fock-expansion-adiabatic}):
\be
\lim_{\epsilon\to 0}\left(e^{-\alpha(\lambda,\omega_R)/(i\epsilon)}\ket{\mathbbm i}\right)=\sum_{n=1}^\infty \check\psi^\text{ad}_{1\to n}(\lambda,\omega_R)\ket{n},
\ee
where $\alpha$ and $\check\psi_{1\to n}^\text{ad}$ are real functions of $\omega_R$ and $\lambda$, consistently with the Gell-Mann and Low theorem. 
Let us give the explicit perturbative expansions of the phase $\alpha$ and of the wave functions. For a generic $\omega_R$, 
\be
\alpha(\lambda,\omega_R)= -\frac{\lambda^2}{\omega }\left(1+2\frac{\lambda^2}{\omega^2}+\frac{38}{3}\frac{\lambda^4}{\omega ^4}+\frac{253}{2}\frac{\lambda^6}{\omega ^6}+\frac{8234}{5}\frac{\lambda^8}{\omega ^8}+25713\frac{\lambda^{10}}{\omega ^{10}}+o(\lambda^{10})\right)
\ee
and
\be
\begin{aligned}
\check\psi^\text{ad}_{1\to1}&=1-\frac{\lambda^2}{\omega ^2}-\frac{19}{2}\frac{\lambda^4}{\omega ^4}-\frac{255}{2}\frac{\lambda^6}{\omega^6}-\frac{16733}{8}\frac{\lambda^8}{\omega ^8}-\frac{315191}{8}\frac{\lambda^{10}}{\omega ^{10}}-\cdots\\
\frac{\check\psi^\text{ad}_{1\to 2}}{\sqrt{2!}}&=-\frac{\lambda}{\omega }-3\frac{\lambda^3}{\omega ^3}-\frac{49}{2}\frac{\lambda^5}{\omega ^5}-\frac{605}{2}\frac{\lambda^7}{\omega ^7}-\frac{38123}{8}\frac{\lambda^9}{\omega ^9}-\frac{709013}{8}\frac{\lambda^{11}}{\omega ^{11}}-\cdots\\
\frac{\check\psi^\text{ad}_{1\to 3}}{\sqrt{3!}}&=\frac{\lambda^2}{\omega ^2}+8\frac{\lambda^4}{\omega ^4}+\frac{193}{2}\frac{\lambda^6}{\omega ^6}+1483\frac{\lambda^8}{\omega ^8}+\frac{215123}{8}\frac{\lambda^{10}}{\omega ^{10}}+\frac{1103795}{2}\frac{\lambda^{12}}{\omega ^{12}}+\cdots
\end{aligned}
\ee
These formulas are not explicit in general: $\omega$ that appears in the right-hand side has to be thought of as the function $\omega=Z_m(\lambda,\omega_R)\times\omega_R$, and the expansion of $Z_m$ depends on the renormalization condition one chooses. Only in the case $\omega_R=\omega$, in which $Z_m=1$, are these expressions fully explicit.

We may check explicitly that unitarity $\sum_n|\check\psi_{1\to n}^\text{ad}|^2=1$ is verified at any order of perturbation theory.


\subsubsection{Physical ground state energy}

Let us work out the relation between the physical and the bare ground state energies, $\omega_P$ and $\omega$ respectively. Then, one will be able to express all wave functions in terms of $\omega_P$.

We can again apply the formula~(\ref{eq:energy-shift-GML}), replacing the asymptotic state $\ket{\Phi_i}$ by $\ket{1}$ and using the expression of $\alpha$ just worked out, setting $\omega_R=\omega$. We find the following relation between $\omega_P$ and $\omega$:
\be
\omega_P=\omega\left(1-2\frac{\lambda^2}{\omega^2}-8\frac{\lambda^4}{\omega^4}-76\frac{\lambda^6}{\omega^6}-1012\frac{\lambda^8}{\omega^8}-16468\frac{\lambda^{10}}{\omega^{10}}-\cdots\right)
\label{eq:omega_P=f(omega)}
\ee

As established earlier (see the discussion just below Eq.~(\ref{eq:energy-shift-GML})), we know that $\alpha(\lambda, \omega_P) = 0$. Substituting the expression for $\alpha$ derived above, with $\omega = Z_m(\lambda, \omega_P) \times \omega_P$, and expanding $Z_m$ as a power series in $\lambda$, this condition can be solved order by order to determine the coefficients of the series defining $Z_m$. This yields
\be
Z_m(\lambda, \omega_P) = 1 + 2 \frac{\lambda^2}{\omega_P^2} + 4 \frac{\lambda^4}{\omega_P^4} + 28 \frac{\lambda^6}{\omega_P^6} + 308 \frac{\lambda^8}{\omega_P^8} + 4372 \frac{\lambda^{10}}{\omega_P^{10}} + \cdots
\label{eq:Zm-expansion}
\ee
One can verify that $\omega = Z_m(\lambda, \omega_P) \times \omega_P$ is indeed the inverse of Eq.~(\ref{eq:omega_P=f(omega)}). 

Equation~(\ref{eq:Zm-expansion}) makes it explicit that only even powers of $\lambda$ appear in the expansion of $Z_m(\lambda, \omega_R = \omega_P)$, thereby justifying, a posteriori, the Ansatz~(\ref{eq:Ansatz-H-0D}).

Finally, let us write down the perturbative expansion of the first few wave functions for $\omega_R=\omega_P$:
\be
\begin{aligned}
\check\psi^\text{ad}_{1\to1}&=1-\frac{\lambda^2}{\omega_P ^2}-\frac{11}{2}\frac{\lambda^4}{\omega_P ^4}-\frac{111}{2}\frac{\lambda^6}{\omega_P^6}-\frac{5997}{8}\frac{\lambda^8}{\omega_P ^8}-\frac{97399}{8}\frac{\lambda^{10}}{\omega_P ^{10}}-\cdots\\
\frac{1}{\sqrt{2!}}\frac{\check\psi^\text{ad}_{1\to 2}}{\check\psi^\text{ad}_{1\to 1}}&=-\frac{\lambda}{\omega_P }-2\frac{\lambda^3}{\omega_P ^3}-{14}\frac{\lambda^5}{\omega_P ^5}-{154}\frac{\lambda^7}{\omega_P ^7}-{2186}\frac{\lambda^9}{\omega_P ^9}-36894\frac{\lambda^{11}}{\omega_P ^{11}}-\cdots\\
\frac{1}{\sqrt{3!}}\frac{\check\psi^\text{ad}_{1\to 3}}{\check\psi^\text{ad}_{1\to 1}}&=\frac{\lambda^2}{\omega_P ^2}+5\frac{\lambda^4}{\omega_P ^4}+47\frac{\lambda^6}{\omega_P ^6}+607\frac{\lambda^8}{\omega_P ^8}+9605\frac{\lambda^{10}}{\omega_P ^{10}}+176169\frac{\lambda^{12}}{\omega_P ^{12}}+\cdots
\end{aligned}
\label{eq:WF-physical-mass}
\ee


\subsubsection{Connection with standard perturbation theory}

In the perturbation theory defined by Eq.~(\ref{eq:standard-PT}) applied to the present problem, only diagrams with multi-particle intermediate states must be taken into account, together with the associated mass counterterm diagrams. The counterterms need to be adjusted in such a way that the energies appearing in the denominators are the eigenvalues of $H$. 

For example, $\widetilde\psi_{1\to 2}$ at order $\lambda^3$ is obtained by adding Eq.~(\ref{eq:1to2-O1}) to~(\ref{eq:1to2-O3-regular}), with $\omega_R$ set to $\omega_P$ in both cases, and $Z_m^{(2)}$ substituted by $2/\omega_P^2$.

Performing this calculation for all orders, the amplitude corresponding to the sum of all $1\to 2$ (resp. $1\to 3$) diagrams evaluated using Eq.~(\ref{eq:standard-PT}) gives exactly the same result as the ratio $\check\psi^\text{ad}_{1\to 2}/\check\psi^\text{ad}_{1\to 1}$ (resp. $\check\psi^\text{ad}_{1\to 3}/\check\psi^\text{ad}_{1\to 1}$) in Eq.~(\ref{eq:WF-physical-mass}), in agreement with the correspondence established in Eq.~(\ref{eq:correspondence}).


\section{Scalar field theory}
\label{sec:phi3}

In this section, we consider the simplest scalar theory with cubic interaction.

\subsection{Definition}

We begin by streamlining the setup of the theory, with the main purpose of fixing notations and conventions. The bare Lagrangian density is
\be
{\cal L}=\frac12\partial_\mu\varphi\,\partial^\mu\varphi-\frac{m^2}{2}\varphi^2-\frac{\lambda}{3!}\varphi^3.
\ee
It will be convenient to work with a renormalized mass $m_R$, defined by $m_R\equiv m/\sqrt{Z_m}$, and a renormalized coupling $\lambda_R$, defined by $\lambda_R\equiv \lambda/Z_\lambda$. In terms of these parameters, the Lagrangian density takes the form
\be
{\cal L}=\frac12\partial_\mu\varphi\,\partial^\mu\varphi-\frac{m_R^2}{2}\varphi^2-\frac{\lambda_R}{3!}\varphi^3+{\cal L}_\text{ct},
\qquad
{\cal L}_\text{ct}=-(Z_m-1)\frac{m_R^2}{2}\varphi^2-(Z_\lambda-1)\frac{\lambda_R}{3!}\varphi^3.
\ee
We keep the space-time dimension \(d\) general, and will specialize to particular values only when needed for comparison with existing calculations. Hence we also need to introduce the arbitrary scale $\mu$, which we use to define the dimensionless renormalized coupling
\[
\bar\lambda_R\equiv \lambda_R\,\mu^{d/2-3}.
\]
The renormalization ``constants'' \(Z_m\) and \(Z_\lambda\) are \textit{a priori} 
functions of \(\bar{\lambda}_R\), and possibly of \(\mu\) and \(m_R\). Finally, we emphasize that we choose not to renormalize the field. Consequently, the wave functions of normalized asymptotic states are, in general, meromorphic functions of the space-time dimension.

We now proceed to the canonical quantization of the theory in a light-cone frame. 
We begin by defining light-cone coordinates. Consider a generic Lorentz vector $V$ in $d$ dimensions with components $(v^0, v^\perp, v^{d-1})$, where $v^\perp=(v^1,v^2,\ldots,v^{d-2})$ denotes a $(d-2)$-dimensional transverse Euclidean vector. We introduce the variables 
\be
v^\pm = \tfrac{1}{\sqrt{2}}\,(v^0 \pm v^{d-1})\,,
\ee
which replace the original components $v^0$ and $v^{d-1}$. For a position $d$-vector $X$, the ``$+$'' component $x^+$ is identified as the light-cone time, while the remaining $d-1$ coordinates are collected in the vector $\vec x\equiv(x^-,x^\perp)$. For a momentum $d$-vector $P$, the ``$-$'' component $p^-$ plays the role of the light-cone energy, conjugate to $x^+$ in the Minkowski product $X\cdot P = x^+p^- + x^-p^+ - x^\perp\cdot p^\perp$, and we define $\vec p\equiv(p^+,p^\perp)$.

The field canonically conjugate to $\varphi$ reads
\be
\frac{\partial{\cal L}}{\partial_+\varphi}=\partial^+\varphi\,.
\ee
Consequently, the Hamiltonian is given by $H={\mathcal H}_{0R}+{\mathcal H}_{1R}$, with
\be
\begin{aligned}
{\mathcal H}_{0R}&=\int d^{d-1}\vec x\left(\frac12\left(\partial^\perp\varphi\right)^2+\frac{m_R^2}{2}\varphi^2\right)\\
\quad\text{and}\quad
{\mathcal H}_{1R}&=\int d^{d-1}\vec x\,\frac{\lambda_R}{3!}\varphi^3+\sum_{j=1}^\infty\bar\lambda_R^{2j}\int d^{d-1}\vec x \left(Z^{(2j)}_m\frac{m^2_R}{2}\varphi^2+Z^{(2j)}_\lambda\frac{\lambda_R}{3!}\varphi^3\right).
\end{aligned}
\label{eq:H-phi}
\ee
We have expanded $Z_m$ and $Z_\lambda$ in powers of $\bar\lambda_R$ as
\be
Z_m=1+\sum_{j=1}^\infty \bar\lambda_R^{2j}\,Z_m^{(2j)},
\quad
Z_\lambda=1+\sum_{j=1}^\infty \bar\lambda_R^{2j}\,Z_\lambda^{(2j)}.
\ee
The series coefficients $Z_m^{(2j)}$ and $Z_\lambda^{(2j)}$ are dimensionless and independent of the coupling.
Each term in ${\cal H}_{1R}$ carries a different power $n$ of $\lambda_R$, which can therefore be used as a label, so that
${\cal H}_{1R} \equiv \sum_{n=1}^\infty {\cal H}_{1R}^{(n)}$.

Introducing the notations
\be
\int \widetilde{dk}\equiv\int_0^{+\infty}\frac{dk^+}{2k^+}\int\frac{d^{d-2}k^\perp}{(2\pi)^{d-1}}
\quad\text{and}\quad
\tilde\delta(\vec k)\equiv 2k^+(2\pi)^{d-1}\delta^{d-1}(\vec k)
\ee
for the boost-invariant integration measure over a single momentum $\vec k$ and the corresponding Dirac distribution respectively, the mode expansion of the field $\varphi$ in the interaction picture reads
\be
\varphi(x)=\int \widetilde{d{k}}\left(a(\vec k)e^{-ikx}+a^\dagger(\vec k)e^{ikx}\right)\,.
\label{eq:mode-expansion-phi}
\ee
The light-cone energy $k^-$ appearing in the phases is related to the momentum $\vec k$ which is integrated over through the mass-shell relation
\be
k^2=2k^+ k^--k^{\perp2}=m_R^2\,.
\ee
It will prove convenient to introduce a notation for the light-cone energy of an on-shell bare particle of 3-momentum $\vec k$:
\be
E_R(\vec k)\equiv \frac{k^{\perp 2}+m_R^2}{2k^+}.
\label{eq:ER-from-mass-shell}
\ee
The subscript ``$R$'' indicates that the mass used in this formula is the renormalized one, $m_R$. We will also make use of the physical light-cone energy $E_P$, defined analogously to $E_R$ but with $m_R$ replaced by the physical mass $m_P$.

The operators $a$ and $a^\dagger$ obey the following commutation relations:
\be
[a(\vec k),a^\dagger(\vec k')]=\tilde\delta(\vec k-\vec{k}'),
\quad [a(\vec k),a(\vec k')]=[a^\dagger(\vec k),a^\dagger(\vec k')]=0\,.
\label{eq:commutation-a}
\ee
The $n$-particle states defined by
\be
\ket{\vec k_1,\cdots,\vec k_n}=a^\dagger(\vec k_1)\cdots a^\dagger(\vec k_n)\ket{0}
\label{eq:states}
\ee
are eigenstates of the free Hamiltonian ${\mathcal H}_{0R}$, with eigenvalues 
\be
\sum_{i=1}^n k_i^-=\sum_{i=1}^n E_R(\vec k_i)\,.
\label{eq:energy-states-n-particles}
\ee
The states~(\ref{eq:states}) obviously correspond to the bare states $\ket{\Phi_j}$. Provided the vacuum state is normalized to unity, the following completeness relation holds in the Fock space:
\be
\sum_j\ket{\Phi_j}\bra{\Phi_j}={\mathbbm I}
\quad\longrightarrow\quad
\sum_n\int \widetilde{dk}_1\cdots \widetilde{d k}_n
\frac{1}{n!}\ket{\vec k_1,\cdots,\vec k_n}\bra{\vec k_1,\cdots,\vec k_n}={\mathbbm I}.
\ee


The evaluation of light-cone wave functions appearing in the Fock state expansions in Eqs.~(\ref{eq:standard-PT}),(\ref{eq:standard-PT-bare}),(\ref{eq:Fock-expansion-i}),(\ref{eq:Fock-expansion-ii}) requires, in addition to the energies~(\ref{eq:energy-states-n-particles}) of $n$-particle intermediate states that enter the denominators, expressions for the matrix elements of the interaction Hamiltonian \( \mathcal{H}_{1R} \). These are readily obtained starting from the expression~(\ref{eq:H-phi}) of \( \mathcal{H}_{1R} \), inserting the mode expansion~(\ref{eq:mode-expansion-phi}), and evaluating its matrix elements between the Fock states~(\ref{eq:states}). We find that all relevant matrix elements are generated by the following elementary ones:
\be
\begin{aligned}
\mel{\vec k_1,\vec k_2}{{\mathcal H}_{1R}}{\vec k}&=
{(2\pi)^{d-1}\delta^{d-1}(\vec k_1+\vec k_2-\vec k)}\,\lambda_R\bigg(
{1}+\sum_{j=1}^\infty 
{Z^{(2j)}_{\lambda} \bar\lambda_R^{2j}}\bigg)\\
\mel{\vec k}{{\mathcal H}_{1R}}{\vec k_1,\vec k_2}&=\mel{\vec k_1,\vec k_2}{{\mathcal H}_{1R}}{\vec k}\\
\mel{\vec k'}{{\mathcal H}_{1R}}{\vec k}&=(2\pi)^{d-1}\delta^{d-1}(\vec k'-\vec k)\sum_{j=1}^\infty
{Z^{(2j)}_m \bar\lambda_R^{2j} m_R^2}
\end{aligned}
\label{eq:matrix-elements-phi3}
\ee
The contribution of order $n$ in $\bar\lambda_R$ to these matrix elements will be  denoted by $\mel{\cdot}{{\cal H}_{1R}^{(n)}}{\cdot}$. In diagrams, the order of a counterterm contribution will be indicated next to the corresponding vertex.

In order not to run into mathematical difficulties with the global momentum conservation between states of fixed momentum in the continuum, we will consider an asymptotic one-particle normalized state of the form
\be
\ket{\phi}\equiv
\int \widetilde{dk}\,\phi(\vec k)\ket{\vec k}
\quad\text{where}\quad
\int \widetilde{dk}\left|\phi(\vec k)\right|^2=1.
\ee
We will denote the corresponding dressed state by
\be
\ket{{\mathbbm i}(\phi)}\equiv U_{IR}^{\text{ad}\,\epsilon}(0,-\infty)\ket{\phi}
=\int \widetilde{dk}\,\phi(\vec k)
\left(U_{IR}^{\text{ad}\,\epsilon}(0,-\infty)\ket{\vec k}\right).
\ee
Our goal will be to compute some of its light-cone wave functions $\psi_{\phi\to\varphi\cdots}^{\text{ad}\,\epsilon}$ to lowest non-trivial order in perturbation theory, using the adiabatic switching prescription. In what follows, we will omit the ``ad'' and ``$\epsilon$'' superscripts on wave functions and dressed states, since no ambiguity should arise.


\subsection{Single-particle wave function}

\begin{figure}
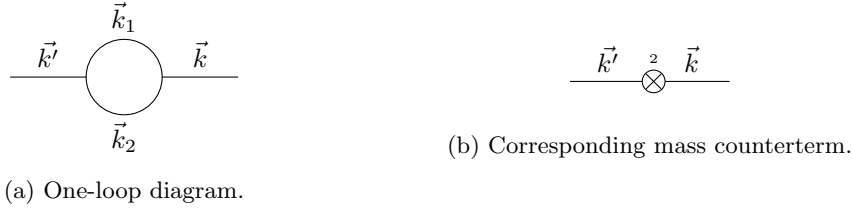

    \begin{center}
    \begin{subfigure}[t]{0.4\textwidth}
    \centerline{\adjustbox{valign=c}{\LConetooneoneloop}}
    \caption{One-loop diagram.}
    \label{subfig:one-to-one-one-loop}
    \end{subfigure}
    \begin{subfigure}[t]{0.4\textwidth}
    \centerline{\adjustbox{valign=c}{\LConetooneoneloopct{$\vec{k'}$}{2}{$\vec{k}$}}}
    \caption{Corresponding mass counterterm.}
    \label{subfig:one-to-one-counterterm}
    \end{subfigure}
    \end{center}
    \caption{Order $\bar\lambda_R^2$ contributions to the $1\to 1$ wave function. The number ``2'' above the counterterm vertex represents its order in powers of $\bar\lambda_R$.}
    \label{fig:one-to-one}
\end{figure}

We now apply light-cone perturbation theory in the adiabatic regularization to compute the probability amplitude to find the dressed state $\ket{\mathbbm{i}(\phi)}$ in a single-particle state of fixed momentum~$\ket{\vec k}$:
\be
\psi_{\phi\to\varphi}(\vec k)\equiv\braket{\vec k}{{\mathbbm i}(\phi)}=\phi(\vec k)+\psi_{\phi\to\varphi}^{(2)}(\vec k)+\cdots.
\ee
The third member introduces the perturbative expansion in the coupling constant, with the superscript on the wave function indicating its order.


\subsubsection{Generic space-time dimension}

Perturbation theory with adiabatic switching, Eq.~(\ref{eq:Fock-expansion-ii}), leads to the following expression for the wave function at order $\bar\lambda_R^2$:
\begin{multline}
\psi_{\phi\to\varphi}^{(2)}(\vec k)=
\int\widetilde{dk'}\phi(\vec{k'})
\,\frac12\int\widetilde{dk_1}\widetilde{dk_2}\frac{\mel{\vec k}{{\cal H}^{(1)}_{1R}}{\vec k_1,\vec k_2}\mel{\vec k_1,\vec k_2}{{\cal H}^{(1)}_{1R}}{\vec k'}}{\left[E_R(\vec{k^{\prime}})-E_R(\vec k)+2i\epsilon\right]\left[E_R(\vec{k^{\prime}})-E_R(\vec{k}_1)-E_R(\vec{k}_2)+i\epsilon\right]}\\
+\int\widetilde{dk'}\phi(\vec{k'})
\frac{\mel{\vec k}{{\cal H}^{(2)}_{1R}}{\vec k'}}{E_R(\vec{k^{\prime}})-E_R(\vec{k})+2i\epsilon}.
\end{multline}
The corresponding diagrams are displayed in Fig.~\ref{subfig:one-to-one-one-loop} (first term) and~\ref{subfig:one-to-one-counterterm} (second term).
We take the matrix elements of the Hamiltonian from Eq.~(\ref{eq:matrix-elements-phi3}). The integration over $k_1$ and $k_2$, together with the symmetry factor $\frac{1}{2}$, arises from the insertion of a sum over a complete set of two-particle states. After a few easy manipulations, we get
\be
\psi_{\phi\to\varphi}^{(2)}(\vec k)=\left(\frac12\frac{\lambda_R^2}{2i\epsilon}\int \frac{\widetilde{dk}_1}{2(k^+-k_1^+)} {\mathbbm 1}_{\{k_1^+\leq k^+\}} \frac{1}{E_R(\vec k)-E_R(\vec k_1)-E_R(\vec k-\vec k_1)+i\epsilon}+\frac{Z_m^{(2)}\bar\lambda_R^2 m_R^2}{2i\epsilon}\right)\frac{\phi(\vec k)}{2k^+}.
\label{eq:psi2phi}
\ee

Let us focus on the first term, corresponding to the first diagram, Fig.~\ref{subfig:one-to-one-one-loop}. We express the light-cone energies in terms of the 3-momenta with the help of the mass-shell relation. Introducing the momentum fraction $z\equiv k_1^+/k^+$, the remaining non-trivial energy denominator reads
\be
E_R(\vec k)-E_R(\vec k_1)-E_R(\vec k-\vec k_1)=-\frac{1}{2k^+}\frac{(k_1^\perp-z k^\perp)^2+[1-z(1-z)]m_R^2}{z(1-z)}.
\label{eq:energy-denominator-2}
\ee
Then, this term can be simplified as
\be
\left.\psi_{\phi\to\varphi}^{(2)}(\vec k)\right|_{\text{Fig.~\ref{subfig:one-to-one-one-loop}}}=-\frac{\bar\lambda_R^2\left(\mu^2\right)^{3-d/2}}{4(4\pi)^{d/2} i\epsilon}\left(\int_0^1 dz \int \frac{d^{d-2} k_1^\perp}{\pi^{d/2-1}}\frac{1}{(k_1^\perp-zk^\perp)^2+m_R^2[1-z(1-z)]-2i\epsilon k^+ z(1-z)}\right)\frac{\phi(\vec k)}{2k^+}.
\label{eq:scalar-1-loop}
\ee
We may shift the integration variable as $k_1^\perp \to k_1^\perp + z k^\perp$, in such a way that the integration over the angles of $k_1^\perp$ becomes trivial. 
We then expand the integrand to first order in $\epsilon$, in such a way that all terms in $\psi_{\phi\to\varphi}^{(2)}$ that survive in the limit $\epsilon \to 0$ are retained. Introducing the master integral
\be
{\cal M}^{(D)}_\nu(V)\equiv \int\frac{d^D \bar k}{\pi^{D/2}}\frac{1}{\left(\bar k^2+V\right)^\nu}=\frac{\Gamma(\nu-D/2)}{\Gamma(\nu)}V^{D/2-\nu},
\label{eq:master-one-loop-body}
\ee
we get
\begin{multline}
\psi_{\phi\to\varphi}^{(2)}(\vec k)
=\frac{\bar\lambda_R^2}{i\epsilon}\left(
\frac{Z_m^{(2)}m_R^2}{2}-\frac{\left(\mu^2\right)^{3-d/2}}{4(4\pi)^{d/2}}\int_0^1 dz\,{\cal M}_1^{(d-2)}[V(z)]\right)\frac{\phi(\vec k)}{2k^+}\\
-\left(\frac{\bar\lambda_R^2(\mu^2)^{3-d/2}}{4(4\pi)^{d/2}}\int_0^1 dz\,z(1-z)\,{\cal M}_2^{(d-2)}[V(z)]\right)
{\phi(\vec k)},
\end{multline}
the argument $V(z)$ of ${\cal M}_{1}$ and ${\cal M}_2$ being defined by $V(z)\equiv m_R^2\left[1-z(1-z)\right]$. Replacing the master integrals by their expressions given in Eq.~(\ref{eq:master-one-loop-body}) and adding the trivial tree-level term, the wave function may be written as
\be
\psi_{\phi\to\varphi}(\vec k)=e^{\alpha(k^+)/(i\epsilon)}\sqrt{Z_\varphi}\,\phi(\vec k)+o(\bar\lambda_R^2)\,,
\label{eq:psi-1-to-2}
\ee
where the phase and modulus read
\be
\begin{aligned}
\alpha(k^+)&=\frac{\bar\lambda_R^2}{2}\frac{m_R^2}{2k^+}\left[
{Z_m^{(2)}}-\left(\frac{m_R^2}{\mu^2}\right)^{d/2-3}\frac{\Gamma(2-d/2)}{2(4\pi)^{d/2}}\int_0^1 dz[1-z(1-z)]^{d/2-2}
\right],\\
\sqrt{Z_\varphi}&=1-{\bar\lambda_R^2}\left(\frac{m_R^2}{\mu^2}\right)^{d/2-3}\frac{\Gamma(3-d/2)}{4(4\pi)^{d/2}}\int_0^1 dz\,z(1-z)[1-z(1-z)]^{d/2-3}.
\end{aligned}
\label{eq:alpha-sqrt(Z)-0}
\ee

We can use the formula~(\ref{eq:energy-shift-GML}) to establish the relationship between the physical mass $m_P$ and the renormalized mass $m_R$. In the light-cone frame, the difference between the energies appearing in the left-hand side of Eq.~(\ref{eq:energy-shift-GML}) reads
\be
E_P(\vec k)-E_R(\vec k)=\frac{k^{\perp2}+m_P^2}{2k^+}-\frac{k^{\perp2}+m_R^2}{2k^+}=\frac{m_P^2-m_R^2}{2k^+}\,.
\ee
Hence Eq.~(\ref{eq:energy-shift-GML}) implies
\be
m_P^2-m_R^2=2k^+\times\bar\lambda_R\frac{\partial\alpha(k^+)}{\partial\bar\lambda_R}=4k^+\alpha(k^+)\times\left[1+O(\bar\lambda_R^2)\right]\,.
\label{eq:energy-shift-GML-phi3-massive}
\ee

The remaining integrals over $z$ in Eqs.~(\ref{eq:alpha-sqrt(Z)-0}) generically amount to ${}_2F_1$ hypergeometric functions. Their expansions near the space-time dimensions $d=4$ (in which the theory is super-renormalizable) and $d=6$ (in which the theory is renormalizable) turn out, however, to be simpler.


\subsubsection{Specializing to four and six dimensions}

\begin{table}[ht]
\centering
\begin{tabular}{lcc}
\hline
 &  $\delta_4\equiv 2-d/2\to 0$ & $\delta_6\equiv 3-d/2\to 0$\\
\hline\\[-1em]
$\sqrt{Z_\varphi}$ & $1-\frac{\bar\lambda_R^2}{4(4\pi)^2}\frac{\mu^2}{m_R^2}\left(\frac{2\pi\sqrt{3}}{9}-1\right)$ & \makecell{$\sqrt{Z_\varphi^{\overline{\text{MS}}}}\times\left[1-\frac{\bar\lambda_R^2}{24(4\pi)^3}\left(\ln\frac{\mu^2}{m_R^2}-\pi\sqrt{3}+\frac{17}{3}\right)\right]$\\
 with $Z_\varphi^{\MSbar}\equiv 1-\frac{\bar\lambda_R^2}{12(4\pi)^3}\left(\frac{1}{\delta_6}-\gamma_E+\ln 4\pi\right)$}\\[1.5em]
$Z_m^\MSbar$ & $1+\frac{\bar\lambda_R^2}{2(4\pi)^2}\frac{\mu^2}{m_{\MSbar}^2}\left(\frac{1}{\delta_4}-\gamma_E+\ln 4\pi\right)$ & $1-\frac{\bar\lambda_R^2}{2(4\pi)^3}\frac56\left(\frac{1}{\delta_6}-\gamma_E+\ln 4\pi\right)$ \\[1.5em]
$\alpha(k^+)$ & $-\frac{\bar\lambda_R^2}{4(4\pi)^2}\frac{\mu^2}{2 k^+}\left(\ln\frac{\mu^2}{m_\MSbar^2}+\frac{\pi\sqrt{3}}{2}-2\right)$ & $\frac{\bar\lambda_R^2}{12(4\pi)^3}\frac{m_\MSbar^2}{2k^+}\left(\frac52\ln\frac{\mu^2}{m_\MSbar^2}-\frac{\pi\sqrt{3}}{2}+\frac{17}{3}\right)$ \\[1.5em]
$m_P^2$ & $m_\MSbar^2-\mu^2\frac{\bar\lambda_R^2}{2(4\pi)^2}\left(\ln\frac{\mu^2}{m_\MSbar^2}+\frac{\pi\sqrt{3}}{2}-2\right)$ & $m_\MSbar^2\left[1-\frac{\bar\lambda_R^2}{6(4\pi)^3}\left(\frac{5}{2}\ln\frac{m_\MSbar^2}{\mu^2}+\frac{\pi\sqrt{3}}{2}-\frac{17}{3}\right)\right]$ \\[1em]
\hline
\end{tabular}
\caption{Explicit expressions, near $d=4$ and $d=6$ and up to order $\bar\lambda_R^2$, for the phase and modulus of the one-particle wave function and for the mass renormalization constant in the $\MSbar$ scheme. The relation between the physical and renormalized masses is also provided.}
\label{tab:self-energy-special-dimensions}
\end{table}

We expand Eq.~(\ref{eq:alpha-sqrt(Z)-0}) in the successive limits $\delta_4 \equiv 2 - d/2 \to 0$ and $\delta_6 \equiv 3 - d/2 \to 0$, retaining only the non-vanishing terms in these limits. We need the following expansions of the integrals over $z$:
\be
\begin{aligned}
\int_0^1 dz\left[1-z(1-z)\right]^{d/2-2}&=
\begin{cases}
1-\delta_4 {\cal J}_0(1)+o(\delta_4)\\
\frac56-\delta_6 {\cal J}_1(1)+o(\delta_6)
\end{cases}\\
\int_0^1 dz\,z(1-z)\left[1-z(1-z)\right]^{d/2-3}&=
\begin{cases}
\frac{2\sqrt{3}\pi}{9}-1+O(\delta_4)\\
\frac16+\delta_6\left[{\cal J}_1(1)-{\cal J}_0(1)\right]+o(\delta_6)
\end{cases}
\end{aligned}
\ee
where we have introduced the notation
\be
{\cal J}_n(x)\equiv\int_0^1 dz\left[1-x z(1-z)\right]^n\ln\left[1-x z(1-z)\right].
\label{eq:def-Jn}
\ee
Explicit expressions for the relevant ${\cal J}$'s are given in Appendix~\ref{appendix:formulas}. The final results for $\sqrt{Z_\varphi}$ and $\alpha$, with the standard $\MSbar$ scheme choice for the mass renormalization constant $Z_m$, are collected in Tab.~\ref{tab:self-energy-special-dimensions}. 
A few comments are in order:
\begin{itemize}
    \item The phase $\alpha$ is \textit{a priori} divergent in both $d=4$ and $d=6$, unless the mass renormalization constant $Z_m$ is suitably adjusted, for example to its value in the $\MSbar$ scheme, $Z_m^\MSbar$. The renormalized mass $m_R\to m_\MSbar$ can then be related to the physical mass using Eq.~(\ref{eq:energy-shift-GML-phi3-massive}). The relation is displayed in Tab.~\ref{tab:self-energy-special-dimensions} for the two considered dimensions.
    
    \item The wave function renormalization constant ${Z_\varphi}$ is finite in $d=4$, but divergent in $d=6$. The divergence may be factorized into the standard $\MSbar$ field renormalization constant, ${Z_\varphi^\MSbar}$.
    
    \item Although we have assumed the theory massive, let us briefly comment on the massless limit in $d=6$. The renormalized and physical masses remain null. But the wave function renormalization constant $Z_\varphi$ diverges logarithmically. Returning to Eq.~(\ref{eq:scalar-1-loop}), we observe that, had we not expanded in a Laurent series in $\epsilon$ before performing the integral over $k_1^\perp$, the adiabatic parameter itself would have regularized the divergence. However, the integration would then have produced logarithms of $\epsilon$, whose interpretation is less clear.
    
\end{itemize}

The expressions we have obtained are consistent with those derived from the covariant formalism; see a standard textbook such as the one in Ref.~\cite{Sterman:1993hfp}. 


\subsection{Two-particle wave function}

We turn to the probability amplitude of our dressed wave packet to be found in a two-particle state of fixed momenta $\vec k_1$ and $\vec k_2$:
\be
\psi_{\phi\to\varphi\varphi}(\vec k_1,\vec k_2)\equiv\braket{\vec k_1,\vec k_2}{{\mathbbm i}(\phi)}=\psi^{(1)}_{\phi\to\varphi\varphi}(\vec k_1,\vec k_2)+\psi^{(3)}_{\phi\to\varphi\varphi}(\vec k_1,\vec k_2)+\cdots
\ee
Again, the last member represents the perturbative expansion in powers of the coupling constant, with superscripts indicating the order of each term. 

The momenta $\vec k_1$ and $\vec k_2$ will be regarded as fixed parameters; to lighten the notation, we shall leave the dependence of the wave functions on them implicit.


\subsubsection{Tree-level diagram}
At order $\lambda_R$, we obtain immediately
\be
\psi^{(1)}_{\phi\to\varphi\varphi}
=\frac{\phi(\vec k_{12})}{2k_{12}^+}
\frac{\lambda_R}{\Delta_R(\vec k_1,\vec k_2)}\,,
\ee
where we introduce the shorthand $\vec k_{12}\equiv \vec k_1+\vec k_2$ for the total momentum of the pair of particles, and 
\be
\Delta_R(\vec k_1,\vec k_2)\equiv E_R(\vec k_{1}+\vec k_{2})-E_R(\vec k_1)-E_R(\vec k_2)
\ee
for the difference between the energy of the asymptotic particle and that of its two-particle Fock state.  
The $i\epsilon$ term in the denominator could be dropped: for fixed, non-exceptional momenta $\vec k_1$ and $\vec k_2$, the limit $\epsilon\to 0$ is clearly nonsingular.  

Next, we introduce the momentum $\vec\kappa_{12}$ with components
\be
\kappa_{12}^+\equiv \frac{k_1^+k_2^+}{k_{12}^+},\qquad 
\kappa_{12}^\perp\equiv \kappa_{12}^+\left(\frac{k_1^\perp}{k_1^+}-\frac{k_2^\perp}{k_2^+}\right).
\ee
If $k_1^+$ and $k_2^+$ are regarded as ``masses'', then $\kappa_{12}^+$ is analogous to the non-relativistic ``reduced mass'' of the pair of particles labeled~1 and~2, while $\kappa_{12}^\perp$ plays the role of their ``relative transverse momentum.'' In terms of these variables, the energy denominator takes the compact form
\be
\Delta_R(\vec k_1,\vec k_2)
=\frac{m_R^2}{2k_{12}^+}
-\frac{\kappa_{12}^{\perp2}+m_R^2}{2\kappa_{12}^+}
=E_R(k_{12}^+,0^\perp)-E_R(\vec\kappa_{12})\,.
\label{eq:def-Delta}
\ee
The analogue of the ``kinetic energy of the center-of-mass'' has manifestly cancelled.

We now turn to radiative corrections, which fall into two categories: self-energy insertions and vertex corrections.


\subsubsection{Self-energy-insertion diagrams}

\begin{figure}
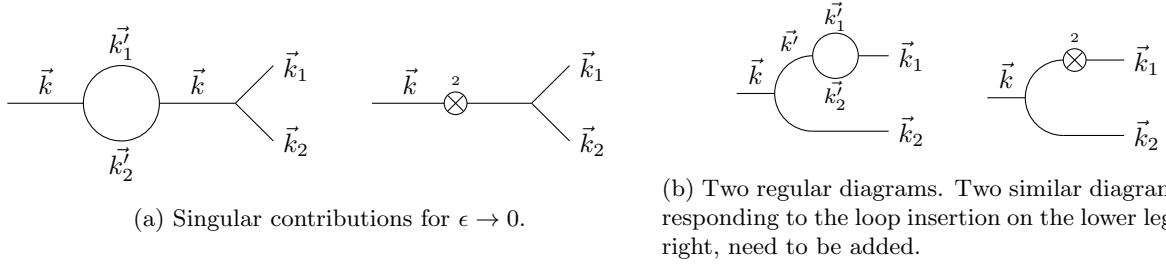

    \begin{center}
    \begin{subfigure}[c]{0.5\textwidth}
    \LConetotwooneloop\quad\quad\LConetotwooneloopct{2}
    \caption{Singular contributions for $\epsilon\to 0$.}
    \label{subfig:singular}
    \end{subfigure}
    \begin{subfigure}[c]{0.45\textwidth}
    \centering\LConetotwobis\quad\quad\LConetotwobisct
    \caption{Two regular diagrams. Two similar diagrams, corresponding to the loop insertion on the lower leg to the right, need to be added.}
    \label{subfig:regular}
    \end{subfigure}
    \end{center}
    \caption{Order $\bar\lambda_R^3$ contributions to the $\phi\to\varphi\varphi$ wave function.}
    \label{fig:1-loop-1-to-2}
\end{figure}

As for self-energy insertions, there are singular and regular contributions in the limit $\epsilon\to 0$, which correspond to the diagrams of Fig.~\ref{subfig:singular} and~\ref{subfig:regular} respectively.

The singular diagrams in Fig.~\ref{subfig:singular} essentially consist of the same factors as those entering the $\phi\to \varphi$ wave function in Eq.~(\ref{eq:psi2phi}), supplemented with factors representing the $\varphi\to\varphi\varphi$ tree-level transition between states of particles of fixed momentum:
\be
\left.\psi_{\phi\to\varphi\varphi}^{(3)}\right|_\text{Fig.~\ref{subfig:singular}}=\frac{\psi_{\phi\to\varphi}^{(2)}(\vec k_{12})}{2k_{12}^+}\frac{\lambda_R}{\Delta_R(\vec k_1,\vec k_2)+3i\epsilon},
\ee
where $\psi_{\phi\to\varphi}^{(2)}$ is the second-order contribution to $\psi_{\phi\to\varphi}$ given in Eq.~(\ref{eq:psi-1-to-2}). Note the term ``$3i\epsilon$'' in the denominator. After replacements and expansion for $\epsilon \to 0$, this reads
\be
\left.\psi_{\phi\to\varphi\varphi}^{(3)}\right|_\text{Fig.~\ref{subfig:singular}}=\lambda_R\left[\left(\frac{\alpha(k_{12}^+)}{i\epsilon}+\sqrt{Z_\varphi}-1\right)\frac{1}{\Delta_R(\vec k_1,\vec k_2)}-\frac{3\alpha(k_{12}^+)}{\Delta_R^2(\vec k_1,\vec k_2)}\right]\frac{\phi(\vec k_{12})}{2k_{12}^+}.
\ee
The terms $\alpha$ and $Z_\varphi$ are given in Eq.~(\ref{eq:alpha-sqrt(Z)-0}).

For the contributions of the regular graphs in Fig.~\ref{subfig:regular}, the two formulations of perturbation theory, the standard one and the adiabatic switching prescription, yield the same expression. We now derive this result algebraically, for bookkeeping purposes, by selecting the appropriate terms in Eq.~(\ref{eq:Fock-expansion-ii}). After simplifying the factors associated with spectator particles at the various interaction times, we obtain
\begin{multline}
\left.\psi_{\phi\to\varphi\varphi}^{(3)}\right|_\text{Fig.~\ref{subfig:regular}}=
\int\widetilde{dk}\,\phi(\vec{k})
\,\frac12\int\widetilde{dk'}\widetilde{dk'_1}\widetilde{dk'_2}\\
\times\frac{\mel{\vec k_1}{{\mathcal H}^{(1)}_{1R}}{\vec k'_1,\vec k'_2 }\mel{\vec k'_1,\vec k'_2}{{\cal H}^{(1)}_{1R}}{\vec{k'}}
\mel{\vec{k'},\vec k_2}{{\cal H}^{(1)}_{1R}}{\vec k}}
{\left[E_R(\vec k)-E_R(\vec k_1)-E_R(\vec k_2)\right]\left[E_R(\vec k)-E_R(\vec k^{\prime}_1)-E_R(\vec k^{\prime}_2)-E_R(\vec k_2)\right]\left[E_R(\vec k)-E_R(\vec{k^{\prime}})-E_R(\vec k_2)\right]}\\
+\int\widetilde{dk}\,\phi(\vec{k})\int\widetilde{dk'}
\frac{\mel{\vec k_1}{{\cal H}^{(2)}_{1R}}{\vec{k'}}\mel{\vec{k'},\vec k_2}{{\cal H}^{(1)}_{1R}}{\vec k}}{\left[E_R(\vec k)-E_R(\vec k_1)-E_R(\vec k_2)\right]\left[E_R(\vec k)-E_R(\vec{k^{\prime}})-E_R(\vec k_2)\right]}\,.
\end{multline}
Replacing the matrix elements and performing the integrals over $\vec k$, $\vec{k'}$ and $\vec{k'_2}$, we arrive at
\begin{multline}
\left.\psi_{\phi\to\varphi\varphi}^{(3)}\right|_\text{Fig.~\ref{subfig:regular}}=
\frac{\lambda_R}{\Delta_R^2(\vec k_1,\vec k_2)}
\bigg(\frac12\int\frac{\widetilde{dk'_1}}{(2\pi)^{d-1} 4k^{+}_1 (k_1^+-k^{\prime+}_{1})}{\mathbbm 1}_{\{k_1^{\prime+}\leq k_1^+\}}
\frac{\lambda_R^2}{\Delta_R(\vec k_1,\vec k_2)+\Delta_R(\vec {k'_1},\vec k_1-\vec{k'_1})}\\
+\bar\lambda_R^2\frac{Z_m^{(2)}}{2k_{12}^{+}}m_R^2\bigg)
\frac{\phi(\vec k_{12})}{2k_{12}^+}.
\end{multline}
Changing the $k_1^{\prime+}$ integration variable to the momentum fraction $z'\equiv k_1^{\prime +}/k_1^{+}$, the energy denominator in the integrand reads, after some algebra,
\be
\Delta_R(\vec k_1,\vec k_2)+\Delta_R(\vec {k'_1},\vec k_1-\vec{k'_1})=-\frac{1}{2k_1^+}\frac{
(k_1^{\prime\perp}-z'k_1^\perp)^2+\left\{1-\left[1-\zeta(\vec k_1,\vec k_2)\right]\times z'(1-z')\right\}m_R^2}{z'(1-z')},
\ee
where
\be
\zeta(\vec k_1,\vec k_2)\equiv -\frac{\Delta_R(\vec k_1,\vec k_2)}{E_R(k_1^+,0^\perp)}.
\label{eq:zeta-denom}
\ee
In the physical region, $\zeta(\vec k_1,\vec k_2)\geq 1$. 

The integral over $k^{\prime\perp}_1$ can now be mapped to the master integral~(\ref{eq:master-one-loop-body}), through an appropriate change of variables. For the sake of writing the result in a compact form, we define
\be
A(\vec k_1,\vec k_2)\equiv\frac{\bar\lambda_R^2}{2} E_R(k_1^+,0^\perp)\Bigg[{Z_m^{(2)}}-\left(\frac{m_R^2}{\mu^2}\right)^{d/2-3}\frac{\Gamma(2-d/2)}{2(4\pi)^{d/2}}\int_0^1 dz'\left\{1-\left[1-\zeta(\vec k_1,\vec k_2)\right]\times z'(1-z')\right\}^{d/2-2}\Bigg].
\label{eq:def-A}
\ee
Comparing this equation with Eq.~(\ref{eq:alpha-sqrt(Z)-0}), we see that $A$ and $\alpha$ would coincide if $\zeta$ were identically zero.
In terms of the function $A$,
\be
\left.\psi_{\phi\to\varphi\varphi}^{(3)}\right|_\text{Fig.~\ref{subfig:regular}}=\frac{\lambda_R}{\Delta_R^2(\vec k_1,\vec k_2)}\times 2{A(\vec k_1,\vec k_2)}\times\frac{\phi(\vec k_{12})}{2k_{12}^+}.
\ee

An expression for the other diagrams, not shown in Fig.~\ref{fig:1-loop-1-to-2} but which can be deduced from those displayed in Fig.~\ref{subfig:regular} by a simple reflexion transformation, is obtained from the previous equation by interchanging $\vec k_1$ and $\vec k_2$. Altogether, the sum of the contributions to the wave function of the tree-level diagram and of the self-energy diagrams reads
\be
\left.\psi_{\phi\to\varphi\varphi}\right|_{\substack{\text{tree level }\\+\text{self-energy}}}=\frac{\phi(\vec k_{12})}{2k_{12}^+}\lambda_R\Bigg(\frac{{\alpha(k_{12}^+)/(i\epsilon)}+\sqrt{Z_\varphi}}{\Delta_R(\vec k_1,\vec k_2)}
+\frac{2{A(\vec k_1,\vec k_2)}+2{A(\vec k_2,\vec k_1)}-3\alpha(k_{12}^+)}{\Delta_R^2(\vec k_1,\vec k_2)}\Bigg).
\label{eq:1-to-2_tree+self-energy-0}
\ee

We can now organize the terms in a more instructive way. From Sec.~\ref{sec:PT}, the singularities as $\epsilon \to 0$ factorize into the phase $e^{\alpha(k_{12}^+)/(i\epsilon)}$, common to all wave functions of the same asymptotic state. After factoring out this phase, we may incorporate some of the terms proportional to $\alpha$ to the energy denominators, taking advantage of the identity~(\ref{eq:energy-shift-GML-phi3-massive}), namely
\be
2\alpha(k^+) = \frac{m_P^2 - m_R^2}{2k^+}+o(\bar\lambda_R^2).
\ee
One easily checks that the following expression matches Eq.~(\ref{eq:1-to-2_tree+self-energy-0}) up to terms of order $\bar\lambda_R^3$:
\be
\left.\psi_{\phi\to\varphi\varphi}\right|_{\substack{\text{tree level }\\+\text{self-energy}}}
=\frac{\phi(\vec k_{12})}{2k_{12}^+}\frac{e^{\alpha(k_{12}^+)/(i\epsilon)} \sqrt{Z_\varphi} \, \lambda_R}{\Delta_P(\vec k_1,\vec k_2) + i\epsilon}
\left( 1 + \frac{2 \left[{A(\vec k_1,\vec k_2)-\alpha(k_1^+)}\right]  + 2 \left[{A(\vec k_2,\vec k_1)-\alpha(k_2^+)}\right] }{\Delta_P(\vec k_1,\vec k_2)} \right).
\label{eq:1-to-2_tree+self-energy}
\ee
To see the equivalence, it is sufficient to expand in powers of $\bar\lambda_R$, recalling that $\alpha$ and $A$ are of order $\bar\lambda_R^2$, keeping only the non-vanishing terms as $\epsilon \to 0$. A few comments are in order:

\begin{itemize}

\item The denominators in Eq.~(\ref{eq:1-to-2_tree+self-energy}) are expressed in terms of the physical mass $m_P$. Actually, the denominator $\Delta_P(\vec k_1,\vec k_2)$ inside the parenthesis could be replaced by $\Delta_R(\vec k_1,\vec k_2)$: it would make no difference at one-loop accuracy.

\item The differences $A-\alpha$ vanish when $\zeta\to 0$. The condition $\zeta=0$ corresponds to the case where the total light-cone energy of the intermediate two-particle state equals that of the external state (i.e.\ the LCPT energy denominator vanishes). Hence there is actually no double pole in Eq.~(\ref{eq:1-to-2_tree+self-energy}) at $\Delta_P(\vec k_1,\vec k_2)=0$: we are effectively left with a single pole corresponding to energy conservation.\footnote{The energy conservation can be realized only at the price of an analytical continuation of transverse momenta to complex vectors.}

\item Depending on the dimension $d$, ultraviolet divergences may still be present, because we have chosen not to renormalize the field $\varphi$. We will investigate them quantitatively below, in the dimension $d=6$ in which the theory is renormalizable.

\end{itemize}

We can also present another version of Eq.~(\ref{eq:1-to-2_tree+self-energy-0}), fully equivalent to the latter at the considered accuracy:
\begin{multline}
\frac{\left.\check{\psi}_{\phi\to\varphi\varphi}\right|_{\substack{\text{tree level }\\+\text{self-energy}}}}{\sqrt{Z_\varphi}}=\frac{\phi(\vec k_{12})}{2k_{12}^+}\Biggl(\frac{\lambda_R}{E_P(\vec k_{12})-E_R(\vec k_{1})-E_R(\vec k_{2})}\\
+\frac{\lambda_R}{\left[E_P(\vec k_{12})-E_R(\vec k_{1})-E_R(\vec k_{2})\right]^2}\left[{{2A(\vec k_1,\vec k_2)}+{2A(\vec k_2,\vec k_1)}}\right]\Biggr).
\label{eq:1-to-2_tree+self-energy-standard}
\end{multline}
We have taken out the singular phase after which we could safely go to the limit $\epsilon\to 0$. This formula would actually be directly obtained from Eq.~(\ref{eq:standard-PT-bare}) projected onto the 2-particle Fock state, expanded at order $\bar\lambda_R^3$, keeping only the tree-level and self-energy diagrams; namely, it coincides with the expression of the wave function calculated in the framework of the standard prescription. The first term corresponds to the tree diagram, the second term corresponds to the sum of the diagrams in Fig.~\ref{subfig:regular}, and the third term is the expression of the diagrams obtained from the latter by swapping the two legs on the right.

Finally, let us choose the renormalized mass to coincide with the physical mass. Then  $\alpha \to 0$, $E_R(\vec k_1) \to E_P(\vec k_{1})$, and $E_R(\vec k_2) \to E_P(\vec k_{2})$. Consequently, Eq.~(\ref{eq:1-to-2_tree+self-energy-standard}) and Eq.~(\ref{eq:1-to-2_tree+self-energy}) divided by $\sqrt{Z_\varphi}$ become manifestly identical once  the limit $\epsilon \to 0$ is taken, which is now regular. This explicitly confirms that the standard and adiabatic prescriptions are fully equivalent for these diagrams: the apparent differences merely reflect a reorganization of self-energy terms. The advantage of the adiabatic switching prescription is that all factors required for properly normalized wave functions emerge directly from the diagrammatic evaluation.

\paragraph{Explicit expressions in $d=6$}

We now expand the wave function for $\delta_6\to 0$, starting from the expression~(\ref{eq:1-to-2_tree+self-energy}). We need the expansion of $A-\alpha$. We first rewrite the difference appearing in one of the terms as
\begin{multline}
{A(\vec k_1,\vec k_2)-\alpha(k_1^+)}=\frac{\bar\lambda_R^2}{4(4\pi)^{3-\delta_6}}E_R(k_1^+,0^\perp)\left(\frac{\mu^2}{m_R^2}\right)^{\delta_6}{\Gamma(\delta_6-1)}\int_0^1 dz'\bigg(\left[1-z'(1-z')\right]^{1-\delta_6}\\
-\left\{1-\left[1-\zeta(\vec k_1,\vec k_2)\right]\times z'(1-z')\right\}^{1-\delta_6}\bigg).
\end{multline}
The non-vanishing terms in the limit $\delta_6\to 0$ can easily be expressed with the help of the ${\cal J}$ integrals defined in Eq.~(\ref{eq:def-Jn}). Dividing by  $\Delta_R(\vec k_1,\vec k_2)$ brings simplifications thanks to the definition~(\ref{eq:zeta-denom}). We then get for one of the terms inside the parenthesis in Eq.~(\ref{eq:1-to-2_tree+self-energy}):
\be
2\frac{A(\vec k_1,\vec k_2)-\alpha(k_1^+)}{\Delta_R(\vec k_1,\vec k_2)}=-\frac{\bar\lambda_R^2}{12(4\pi)^3}
\left(\frac{1}{\delta_6}-\gamma_E+\ln 4\pi+\ln\frac{\mu^2}{m_R^2}+1+\frac{6}{\zeta(\vec k_1,\vec k_2)}\left\{{\cal J}_1(1)-{\cal J}_1[1-\zeta(\vec k_1,\vec k_2)]\right\}\right).
\label{eq:A-alpha-final}
\ee
The other non-trivial term in Eq.~(\ref{eq:1-to-2_tree+self-energy}) is deduced from Eq~(\ref{eq:A-alpha-final}) by substituting $\vec k_1\leftrightarrow\vec k_2$. Noticing that the singularity when $\delta_6\to 0$ may be extracted with the help of two factors $Z_\varphi^\MSbar$, we may write the full expression of the self-energy contributions to the wave function as
\begin{multline}
\left.\psi_{\phi\to\varphi\varphi}\right|_{\substack{\text{tree level }\\+\text{self-energy}}}=\frac{\phi(\vec k_{12})}{2k_{12}^+}
\frac{e^{\alpha(k_{12}^+)/(i\epsilon)} \sqrt{Z_\varphi} \, \lambda_R}{\Delta_P(\vec k_1,\vec k_2) + i\epsilon}\\
\times\left(Z_\varphi^\MSbar\right)^2
\left[1-\frac{\bar\lambda_R^2}{2(4\pi)^3}\left(\frac13+\frac{1}{3}\ln\frac{\mu^2}{m_R^2}+\frac{{\cal J}_1(1)-{\cal J}_1[1-\zeta(\vec k_1,\vec k_2)]}{\zeta(\vec k_1,\vec k_2)}+\frac{{\cal J}_1(1)-{\cal J}_1[1-\zeta(\vec k_2,\vec k_1)]}{\zeta(\vec k_2,\vec k_1)}\right)\right].
\label{eq:final-self-energy}
\end{multline}
To complete the calculation, one substitutes the ${\cal J}$ functions with their explicit expressions given in Appendix~\ref{appendix:formulas}, after having replaced their arguments using Eqs.~(\ref{eq:zeta-denom}),(\ref{eq:def-Delta}),(\ref{eq:ER-from-mass-shell}). The final expression in terms of the kinematical variables $\vec k_1$ and $\vec k_2$ is not particularly illuminating. Let us instead highlight a few facts:

\begin{itemize}

\item The ultraviolet singularities coincide with those of the factor 
$\sqrt{Z_\varphi^{\overline{\rm MS}}} \times \left(Z_\varphi^{\overline{\rm MS}}\right)^{2}$, i.e., the renormalization factor associated with the product of a single-particle wave function (that of the asymptotic particle) and two propagators (propagating the two particles resulting from the fluctuation up to light-cone time~0). This behavior was also observed in Ref.~\cite{Lappi:2016oup} in a similar two-particle wave function calculation, but in the context of gauge theories.

\item If one analytically continues the transverse momenta so that the $\zeta$'s tend to zero, which corresponds to making the intermediate-state light-cone energy equal to the external one, the second line on the right-hand side of the previous equation reduces to $\left(Z_\varphi\right)^2$, where the expression of $Z_\varphi$ is given in Tab.~\ref{tab:self-energy-special-dimensions}. Physically, this reflects the fact that when energy conservation is enforced, the lifetime of the particle pair becomes infinite, and thus, the fluctuations of each of these particles becomes part of the dressing of an asymptotic state.

\item The limit $m\to 0$ (implying $m_R,m_P\to 0$ in space-time dimension $d=6$) of the ratio of Eq.~(\ref{eq:final-self-energy}) to the $\phi\to\varphi$ wave function is regular. Physically, this is because the loop on the outgoing legs can only occur during a finite time, due to the light-cone energy difference between the two-particle intermediate state and the external particle: the inverse of this difference sets a soft upper bound on the lifetime of the two-particle state, and thus on the time interval during which radiative corrections to the propagation of each particle can take place. Hence, infrared divergences associated to these fluctuations cannot occur.

One finds
\be
\left.\frac{\left.{\psi}_{\phi\to\varphi\varphi}\right|_{\substack{\text{tree level }\\+\text{self-energy}}}}{\psi^{(1)}_{\phi\to\varphi\varphi}}\right|_{m\to 0}=
\left(Z_\varphi^\MSbar\right)^2\left[1+\frac{\bar\lambda_R^2}{6(4\pi)^3}\left(\ln\frac{\kappa_{12}^{\perp2}}{\mu^2}+\frac12\ln\frac{k_{12}^+}{\kappa_{12}^+}-\frac83\right)\right].
\ee

\end{itemize}


\subsubsection{Vertex corrections}
\label{subsec:vertex-correction}

\begin{figure}
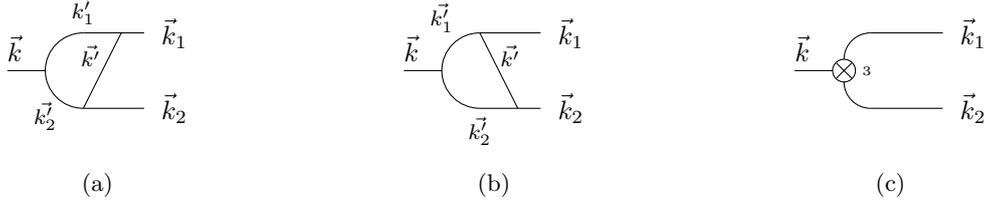

    \begin{center}
    \begin{subfigure}[b]{.3\textwidth}
        \centerline{\LCtrianglebis}
        \caption{}
        \label{subfig:1to2-bare1}
    \end{subfigure}
    \begin{subfigure}[b]{.3\textwidth}
        \centerline{\LCtriangle}
        \caption{}
        \label{subfig:1to2-bare2}
    \end{subfigure}
    \begin{subfigure}[b]{.3\textwidth}
        \centerline\LCtrianglect
        \caption{}
        \label{subfig:1to2-ct}
    \end{subfigure}
    \end{center}
    \caption{One-loop vertex correction diagrams.}
    \label{fig:triangle-graphs}
\end{figure}

As for the one-loop corrections to the vertex, their computation is identical whether one employs the standard formulation or the adiabatic switching prescription of perturbation theory. For completeness, we nevertheless display the calculation here.

We begin with the diagram shown in Fig.~\ref{subfig:1to2-bare1}, where a particular assignment of momenta is indicated. Its contribution to the two-particle wave function reads
\begin{multline}
\left.\psi^{(3)}_{\phi\to\varphi\varphi}\right|_\text{Fig.~\ref{subfig:1to2-bare1}}=\int\widetilde{dk}\,\phi(\vec{k})
\int\widetilde{dk'}\widetilde{dk'_1}\widetilde{dk'_2}\frac{1}{E_R(\vec k)-E_R(\vec k_1)-E_R(\vec k_2)}\\
\times\frac{\mel{\vec{k}_1}{{\mathcal H}^{(1)}_{1R}}{\vec k'_1,\vec {k'}}\mel{\vec{k'},\vec k_2}{{\cal H}^{(1)}_{1R}}{\vec k'_2}
\mel{\vec k'_1,\vec k'_2}{{\cal H}^{(1)}_{1R}}{\vec k}}
{\left[E_R(\vec k)-E_R(\vec k^{\prime}_1)-E_R(\vec k^{\prime})-E_R(\vec k_2)\right]\left[E_R(\vec k)-E_R(\vec k_1^{\prime})-E_R(\vec k_2^{\prime})\right]}\,.
\end{multline}
Note that there is no combinatorial factor from the insertion of the intermediate states because the momenta label all particles unambiguously. 

After replacing the matrix elements of the Hamiltonian, whose momentum-conservation factors appear as delta functions, we can carry out the integrations over $\vec{k}$, $\vec{k'}$, and $\vec{k}'_2$, which then become trivial. The energy denominators can subsequently be expressed in terms of the momentum components. Among them, the two that depend on the remaining integration variables $(k_1^{\prime+} \equiv z' k_{12}^+,\, k_1^{\prime\perp})$, namely, the last two, take the form
\be
\begin{aligned}
D_1&\equiv E(\vec k_{12})-E(\vec k_1^{\prime})-E(\vec k_{12}-\vec k_1^{\prime})=-\frac{1}{2k_{12}^+}\frac{\left(k^{\prime\perp}_1-z'k_{12}^\perp\right)^2+\left[1-z'(1-z')\right]m_R^2}{z'(1-z')}\\
D_2&\equiv E(\vec k_{12})-E(\vec k^{\prime}_1)-E(\vec k_1-\vec k_1^{\prime})-E(\vec k_2)\\
&=-\frac{1}{2k_{12}^+}\frac{
\left(k^{\prime\perp}_1-{z'}k_1^\perp k_{12}^+/k_1^+\right)^2+{z'(1-z'k_{12}^+/k_1^+)}\left(k_{12}^{\perp2}k_{12}^+/\kappa_{12}^+\right)+{(1-z')(1+z'k_{12}^+/k_2^+)}m_R^2}{z'(1-z'k_{12}^+/k_1^+)}\,.
\end{aligned}
\ee
We group them introducing an integral over a Feynman parameter, that we implement as follows:
\be
\frac{1}{D_1D_2}=\int_0^1 {ds}\frac{z^{\prime2}(1-z')(1-z'k_{12}^+/k_1^+)}{\left[(1-s)z'(1-z')D_1+s z'(1-z'k_{12}^+/k_1^+)D_2\right]^2}.
\ee
After a few straightforward manipulations, we get
\be
\left.\psi^{(3)}_{\phi\to\varphi\varphi}\right|_\text{Fig.~\ref{subfig:1to2-bare1}}=\psi^{(1)}_{\phi\to\varphi\varphi}\times
\frac{\lambda_R^2}{(4\pi)^{d/2}}\frac{k_{12}^+}{k_1^+}
\int_0^1 ds\int_0^{k_1^+/k_{12}^+} dz'\,{z'}{\cal M}_2^{d-2}[V(z',s)]\,,
\ee
where the argument $V$ of the master integral $\cal M$ reads
\be
V(z',s)\equiv s{z'(1-z')}\left(1-s\frac{z'}{1-z'}\frac{k_2^+}{k_1^+}\right)\frac{k_{12}^+{\kappa_{12}^{\perp 2}}}{\kappa_{12}^+}+\left[1-z'(1-z')+sz'\left((1-z')\frac{k_{12}^+}{k_2^+}-z'\right)\right]m_R^2\,.
\ee
Replacing $\cal M$ by its expression~(\ref{eq:master-one-loop-body}), we get
\begin{multline}
\frac{\left.\psi^{(3)}_{\phi\to\varphi\varphi}\right|_\text{Fig.~\ref{subfig:1to2-bare1}}}{\psi^{(1)}_{\phi\to\varphi\varphi}}=\frac{\bar\lambda_R^2}{(4\pi)^{d/2}}\Gamma(\delta_6)\frac{k_{12}^+}{k_1^+}
\int_0^1 ds \int_0^{k_1^+/k_{12}^+} dz'{z'}\Bigg\{s{z'(1-z')}\left(1-s\frac{z'}{1-z'}\frac{k_2^+}{k_1^+}\right)\frac{k_{12}^+}{\kappa_{12}^+}\frac{\kappa_{12}^{\perp 2}}{\mu^2}\\
+\left[1-z'(1-z')+sz'\left((1-z')\frac{k_{12}^+}{k_2^+}-z'\right)\right]\frac{m_R^2}{\mu^2}\Bigg\}^{-\delta_6}.
\label{eq:1to2-vertex1-final}
\end{multline}
In general dimension $d$, one cannot perform the integrals in terms of elementary functions.

As for the remaining diagrams in Fig.~\ref{fig:triangle-graphs}:
\begin{itemize}
\item The expression of the diagram of Fig.~\ref{subfig:1to2-bare2} is straightforward to deduce from that of the diagram of Fig.~\ref{subfig:1to2-bare1}:
\be
\left.\psi_{\phi\to\varphi\varphi}^{(3)}(\vec k_1,\vec k_2)\right|_\text{Fig.~\ref{subfig:1to2-bare2}}=\left.\psi^{(3)}_{\phi\to\varphi\varphi}(\vec k_2,\vec k_1)\right|_\text{Fig.~\ref{subfig:1to2-bare1}}.
\ee
\item The counterterm diagram of Fig.~\ref{subfig:1to2-ct} reads
\be
{\left.\psi^{(3)}_{\phi\to\varphi\varphi}\right|_\text{Fig.~\ref{subfig:1to2-ct}}}=\int\widetilde{dk}\,\phi(\vec k)\frac{\mel{\vec k_1,\vec k_2}{{\mathcal H}_{1 R}^{(3)}}{\vec k}}{E_R(\vec k)-E_R(\vec k_1)-E_R(\vec k_2)}=
{\psi^{(1)}_{\phi\to\varphi\varphi}}\times Z_\lambda^{(2)}\bar\lambda_R^2\,.
\ee
\end{itemize}
We shall work out an appropriate choice for the constant $Z_\lambda^{(2)}$ in the dimension 6 in which the theory is renormalizable.


\paragraph{Explicit expressions in $d=6$}

The divergent terms are straightforward to calculate: it is enough to write $\Gamma(\delta_6)=\Gamma(1+\delta_6)/\delta_6$, and, subsequently, set $d=6$ in all factors in Eq.~(\ref{eq:1to2-vertex1-final}), except $1/\delta_6$. We get
\be
{\left.\psi^{(3)}_{\phi\to\varphi\varphi}\right|_{\substack{\text{Fig.~\ref{subfig:1to2-bare1}}\\ \text{UV singular}}}}={\psi^{(1)}_{\phi\to\varphi\varphi}}\times\frac{\bar\lambda_R^2}{2(4\pi)^3}\frac{1}{\delta_6}\times \frac{k_1^+}{k_{12}^+}.
\ee
The amplitude corresponding to the diagram of Fig.~\ref{subfig:1to2-bare2} has the same expression, up to the substitution $k_1^+\to k_2^+$. Hence, the divergent term in the renormalization constant has to be set to the following value:
\be
\left. Z_\lambda^{(2)}\right|_{\text{UV singular}}=-\frac{1}{2(4\pi)^3}\frac{1}{\delta_6}.
\ee
It coincides with that obtained in covariant calculations~\cite{Sterman:1993hfp}.

There is no simple expression for the finite terms in the wave function in the massive theory. But for $m_R=0$, Eq.~(\ref{eq:1to2-vertex1-final}) reduces, in the limit $\delta_6\to 0$, to
\be
\left.\frac{\left.\psi^{(3)}_{\phi\to\varphi\varphi}\right|_\text{Fig.~\ref{subfig:1to2-bare1}}}{\psi^{(1)}_{\phi\to\varphi\varphi}}\right|_{m\to 0}=\frac{\bar\lambda_R^2}{(4\pi)^{3}}\bigg[\frac{z}{2}\left(\frac{1}{\delta_6}-\gamma_E+\ln 4\pi-\ln\frac{\kappa_{12}^{\perp2} k_{12}^+}{\mu^2\kappa_{12}^+}\right)-f\left(\frac{k_1^+}{k_{12}^+}\right)\Bigg],
\ee
where
\be
\begin{aligned}
f(z)&\equiv\int_0^1 ds\int_0^z dz'\frac{z'}{z}\left\{\ln s+\ln z'+\ln\left[1-z'\left(1+s\frac{1-z}{z}\right)\right]\right\}\\
&=\frac{z}{2}\ln z-\frac{1-z}{2}\ln(1-z)-\frac32 z\,.
\end{aligned}
\ee
All integrals appearing in \(f\) could be evaluated exactly: indeed, after a few elementary manipulations, consisting essentially of shifts of the integration variables, they boil down to a sum of integrals of the form~(\ref{eq:elem-log-int}). The final result reads
\be
\left.\frac{\left.\psi^{(3)}_{\phi\to\varphi\varphi}\right|_\text{Fig.~\ref{subfig:1to2-bare1}}}{\psi^{(1)}_{\phi\to\varphi\varphi}}\right|_{m\to 0}=\frac{\bar\lambda_R^2}{2(4\pi)^3}\bigg[z\left(\frac{1}{\delta_6}-\gamma_E+\ln 4\pi+3-\ln\frac{\kappa_{12}^{\perp2} k_{12}^+}{\mu^2\kappa_{12}^+}\right)
-\frac{k_1^+}{k_{12}^+}\ln \frac{k_1^+}{k_{12}^+}+\frac{k_2^+}{k_{12}^+}\ln\frac{k_2^+}{k_{12}^+}
\bigg].
\ee
The diagram of Fig.~\ref{subfig:1to2-bare2} has the same expression, up to the substitution $\vec k_1\leftrightarrow\vec k_2$. The sum of the diagrams~\ref{subfig:1to2-bare1} and~\ref{subfig:1to2-bare2} then has a very simple expression, since the last two terms in the previous equation cancel between the two diagrams. Setting the renormalization constant to its $\MSbar$ value, namely
\be
Z_\lambda=Z_\lambda^\MSbar\equiv 1-\frac{\bar\lambda_R^2}{2(4\pi)^3}\left(\frac{1}{\delta_6}-\gamma_E+\ln 4\pi
\right),
\ee
the full one-loop vertex correction contribution to the wave function, namely the sum of the contributions of the three diagrams shown in Fig.~\ref{fig:triangle-graphs}, reads
\be
\left.\frac{\left.\psi^{(3)}_{\phi\to\varphi\varphi}\right|_{\substack{\text{vertex corrections}\\\text{$\MSbar$}}}}{\psi^{(1)}_{\phi\to\varphi\varphi}}\right|_{m\to 0}=-\frac{\bar\lambda_R^2}{2(4\pi)^3}\left(\ln\frac{\kappa_{12}^{\perp 2}}{\mu^2}+\ln\frac{k_{12}^+}{\kappa_{12}^+}-3\right).
\ee


\section{Conclusion}
\label{sec:conclusion}
The starting point of light-cone perturbation theory is the (light-cone) time-evolution operator in the interaction picture. In this work, we have compared two ways of regularizing this operator: the standard prescription, which underlies most modern calculations of light-cone wave functions, and adiabatic switching.  

Both prescriptions can be associated with a physical picture. The one underlying the standard prescription is arguably more physically realistic, as it corresponds to preparing a particle at some finite time and allowing it to propagate with the full physical Hamiltonian. From a technical perspective, this formalism is directly related to a resolvent of the Schrödinger equation. As such, it has become a well-established and widely used framework for computing wave functions in light-cone perturbation theory at one-loop accuracy. By contrast, adiabatic switching may appear more artificial, since it amounts to turning off the interaction at asymptotic times. Nevertheless, this idea is closely connected to scattering from an external potential in non-relativistic quantum mechanics, where the bare particles are physical -- particularly when the potential is localized in a bounded region of space. At the same time, adiabatic switching offers the advantage of enabling fully diagrammatic calculations of properly normalized wave functions, which can be conceptually significant in certain contexts. For example, the correct implementation of the adiabatic switching prescription played a crucial role in the diagrammatic derivation of the color dipole model in Ref.~\cite{Chen:1995pa}. We also note the recent proposal for constructing an infrared-finite $S$-matrix in Ref.~\cite{Hannesdottir:2019opa}, where adiabatic switching is implicitly employed to obtain perturbative expansions of matrix elements of asymptotic Møller operators.\footnote{Compare Eq.~(58) in Ref.~\cite{Hannesdottir:2019opa} with our Eq.~(\ref{eq:Fock-expansion-i}), and see Sec.~III B therein for an example of a calculation in $\varphi^3$ theory. Note that in Ref.~\cite{Hannesdottir:2019opa} the theory is quantized in a Lorentz frame.}

To illustrate the formalism, we restricted our attention to scalar theories so as not to dilute the discussion. We first analyzed a simple quantum-mechanical cubic anharmonic oscillator, for which calculations can be carried out straightforwardly to arbitrary orders in perturbation theory. We then turned to massive $\varphi^3$ field theory, where the only additional complication compared to the quantum-mechanical model arises from the familiar issue of ultraviolet divergences, which the formalism successfully accommodates. Our work clarifies, at both the conceptual and computational levels, the differences between the two prescriptions and demonstrates explicitly how adiabatic switching can be used as a fully diagrammatic tool for normalized wave functions.  

As discussed, the relative simplicity of the massive scalar field theory does not carry over to the massless case. The adiabatic parameter~$\epsilon$ regularizes collinear singularities by cutting off radiation at early or late times, but in doing so introduces logarithmic singularities in~$\epsilon$, which is somewhat awkward. A natural outlook is therefore to develop a systematic treatment of infrared singularities within this framework. This would make it possible to perform one- and two-loop calculations in the massless scalar theory and compare them with existing results obtained in covariant perturbation theory, eventually paving the way toward extending the formalism to gauge theories such as QED and QCD.


\section*{Acknowledgements} 

We are grateful to A.-K. Angelopoulou, F. Gelis, A.~H. Mueller and U. Reinosa for helpful discussions at various stages in the completion of this work. We also thank L.~Szymanowski for his interest and for his careful reading of the manuscript.


\appendix

\section{Useful formulas}
\label{appendix:formulas}

We give here explicit expressions for the integral ${\cal J}_n(x )$ defined in Eq.~(\ref{eq:def-Jn}), for relevant values of the parameters. 

For any real number $x <4$,
\be
\begin{aligned}
{\cal J}_0(x )=&\int_0^1 dz\, \ln\left[1-x  z(1-z)\right]=-2+\begin{cases}
2\sqrt{\frac{4}{x }-1}\,\arctan\frac{1}{\sqrt{4/x -1}} &\text{for $0<x <4$}\\
\sqrt{1-\frac{4}{x }}\,\ln\left[1-\frac{x }{2}\left(1+\sqrt{1-\frac{4}{x }}\right)\right] & \text{for $x <0$}
\end{cases}\\
{\cal J}_1(x )=&\int_0^1 dz\, \left[1-x  z(1-z)\right]\ln\left[1- x  z(1-z)\right]=\\
&\hspace{4cm}\frac{5 x  }{18}-\frac{4}{3}+\frac{ x  }{6}\times
\begin{cases}
2\left(\frac{4}{ x  }-1\right)^{3/2}\,\arctan\frac{1}{\sqrt{4/ x  -1}}& \text{for $0< x  <4$}\\
\left(1-\frac{4}{ x  }\right)^{3/2}\,\ln\left[1-\frac{ x  }{2}\left(1+\sqrt{1-\frac{4}{ x  }}\right)\right] & \text{for $ x  <0$}
\end{cases}
\end{aligned}
\ee
(These integrals would need to be defined for $ x  \geq 4$ because of the presence of a branch cut). The following particular cases prove useful:
\be
\begin{aligned}
{\cal J}_0(1)=&\int_0^1 dz\, \ln\left[1-z(1-z)\right]=\frac{\sqrt{3}}{3}\pi-2\,,\\
{\cal J}_1(1)=&\int_0^1 dz\, \left[1-z(1-z)\right]\ln\left[1-z(1-z)\right]=\frac16\left({\sqrt{3}}\,\pi-\frac{19}{3}\right).
\end{aligned}
\label{eq:integrals-over-z}
\ee

An elementary way to evaluate these integrals is to first factorize the argument 
\(1 - x\,z(1-z)\) of the logarithm, either over \(\mathbb{R}\) (for \(x \le 0\)) or over \(\mathbb{C}\) (for \(0 < x < 4\)). After suitable shifts of the integration variable, one obtains a sum of integrals, all of which follow from the indefinite integral
\be
\int dt\, t^{n-1} \ln t 
= \frac{t^n}{n} \left( \ln t - \frac{1}{n} \right) + C.
\label{eq:elem-log-int}
\ee



\input biblio.bbl

\end{document}

%% file: graphics.tikz
\def\leg{\scalebox{1.}{\begin{tikzpicture}[baseline={([yshift=-2.5]current bounding box.center)}]
  \begin{feynman}
  \vertex(a1) at (0,0);
  \vertex(a11) at (0.5,0);
    \diagram* {
    (a1)--(a11);
    };
  \end{feynman}
\end{tikzpicture}}}

\def\LConetooneoneloop{\scalebox{1.}{\begin{tikzpicture}[baseline={(current bounding box.center)}]
  \begin{feynman}
  \draw[-] (-0.5,0) to (0.5,0);
  \node at (0,0.3) {$\vec{k'}$};
  \draw(1,0) circle [radius=0.5];
  \draw[-](1.5,0) to (2.5,0);
  \node at (1,0.8) {$\vec{k}_1$};
  \node at (1,-0.8) {$\vec{k}_2$};
  \node at (2,0.3) {$\vec{k}$};
  \end{feynman}
\end{tikzpicture}}}

\def\LConetooneoneloopct#1#2#3{\scalebox{1.}{\begin{tikzpicture}[baseline={(current bounding box.center)}]
  \begin{feynman}
  \draw[-](-0.1,0) to (0.85,0);
  \node at (0.4,0.3) {#1};
  \draw(1,0) circle [radius=0.15];
  \draw(0.9,-0.1) to (1.1,0.1);
  \draw(0.9,0.1) to (1.1,-0.1);
  \node at (1, 0.3) {\tiny #2};
  \node at (1,-0.3) {\phantom{\tiny #2}};
  \draw[-](1.15,0) to (2,0);
  \node at (1.5,0.3) {#3};
  \end{feynman}
\end{tikzpicture}}}

\def\LConetotwooneloop{\scalebox{1.}{\begin{tikzpicture}[baseline={([yshift=-2.5]current bounding box.center)}]
  \begin{feynman}
  \draw[-] (-0.5,0) to (0.5,0);
  \node at (0,0.3) {$\vec{k}$};
  \draw(1,0) circle [radius=0.5];
  \draw[-](1.5,0) to (2.5,0);
  \node at (1,0.8) {$\vec{k'_1}$};
  \node at (1,-0.8) {$\vec{k'_2}$};
  \node at (2,0.3) {$\vec{k}$};
  \draw[-] (2.5,0) to (3,0.5);
  \node at (3.3,0.5) {$\vec{k}_1$};
  \draw[-] (2.5,0) to (3,-0.5);
  \node at (3.3,-0.5) {$\vec{k}_2$};
  \end{feynman}
\end{tikzpicture}}}

\def\LConetotwooneloopct#1{\scalebox{1.}{\begin{tikzpicture}[baseline={([yshift=-2.5]current bounding box.center)}]
  \begin{feynman}
  \draw[-](-0.1,0) to (0.85,0);
  \node at (0.4,0.3) {$\vec{k}$};
  \draw(1,0) circle [radius=0.15];
  \draw(0.9,-0.1) to (1.1,0.1);
  \draw(0.9,0.1) to (1.1,-0.1);
  \node at (1, 0.3) {\tiny #1};
  \node at (1,-0.3) {\phantom{\tiny #1}};
  \draw[-](1.15,0) to (2,0);
  \draw[-] (2,0) to (2.5,0.5);
  \node at (2.8,0.5) {$\vec{k}_1$};
  \draw[-] (2,0) to (2.5,-0.5);
  \node at (2.8,-0.5) {$\vec{k}_2$};
  \end{feynman}
\end{tikzpicture}}}

\def\LConetotwobis{\scalebox{1.}{\begin{tikzpicture}[baseline={([yshift=-2.5]current bounding box.center)}]
  \begin{feynman}
    \node at (-0.75,-0.2) {$\vec{k}$};
    \draw(-1.,-0.5) to (-0.5,-0.5);
    \draw(0.,0) arc (90:270:0.5);
    \node at (-0.3,0.2) {\footnotesize $\vec{k'}$};
    \draw(0.3,0.) circle [radius=0.3];
    \node at (0.3,0.52) {\footnotesize $\vec{k'_1}$};
    \node at (0.3,-0.53) {\footnotesize $\vec{k'_2}$};
    \draw(0.6,0) to (1.,0);
    \draw(0.,-1) to (1.,-1);
    \node at (1.3,0) {$\vec{k}_1$};
    \node at (1.3,-1) {$\vec{k}_2$};
    \end{feynman}
\end{tikzpicture}}}

\def\LConetotwobisct{\scalebox{1.}{\begin{tikzpicture}[baseline={([yshift=-2.5]current bounding box.center)}]
  \begin{feynman}
  \node at (-0.25,-0.2) {$\vec{k}$};
    \draw(-0.5,-0.5) to (0.,-0.5);
    \draw(0.5,0) arc (90:270:0.5);
    \draw(0.65,0.) circle [radius=0.15];
    \draw(0.65-0.1,0.-0.1) to (0.65+0.1,0.+0.1);
    \draw(0.65-0.1,0.+0.1) to (0.65+0.1,0.-0.1);
    \node at (0.65, 0.+0.3) {\tiny 2};
    \node at (0.3,0.6) {\phantom{$\vec{k'_1}$}};
    \draw(0.8,0) to (1.3,0);
    \draw(0.5,-1) to (1.3,-1);
    \node at (1.6,0) {$\vec{k}_1$};
    \node at (1.6,-1) {$\vec{k}_2$};
    \end{feynman}
\end{tikzpicture}}}

\def\LCtriangle{\scalebox{1.}{\begin{tikzpicture}[baseline={([yshift=-2.5]current bounding box.center)}]
  \begin{feynman}
  \node at (0.1,-0.2) {$\vec{k}$};
    \node at (1,0.3) {\footnotesize \phantom{$\vec{k_1'}$}};
  \node at (0.5,0.2) {\footnotesize $\vec{k_1'}$};
  \node at (1,-1.3) {\footnotesize $\vec{k_2'}$};
  \node at (1.4,-.3) {\footnotesize $\vec{k'}$};
  \draw[-](0,-0.5) to (0.5,-0.5);
  \draw(1,0) arc (90:270:0.5);
  \draw[-] (1,0) to (1.5,-1);
  \draw[-] (1,0) to (1.8,0.0);
  \draw[-] (1,-1) to (1.8,-1);
  \node at (2.2,0) {$\vec k_1$};
  \node at (2.2,-1) {$\vec k_2$};
  \end{feynman}
\end{tikzpicture}}}

\def\LCtrianglebis{\scalebox{1.}{\begin{tikzpicture}[baseline={([yshift=-2.5]current bounding box.center)}]
  \begin{feynman}
  \node at (0.1,-0.2) {$\vec{k}$};
  \node at (1,0.3) {\footnotesize $\vec{k_1'}$};
  \node at (0.5,-1.1) {\footnotesize $\vec{k_2'}$};
  \node at (1.1,-.3) {\footnotesize $\vec{k'}$};
    \node at (1,-1.3) {\footnotesize \phantom{$\vec{k_2'}$}};
  \draw[-](0,-0.5) to (0.5,-0.5);
  \draw(1,0) arc (90:270:0.5);
  \draw[-] (1.5,0) to (1,-1);
  \draw[-] (1,0) to (1.8,0.0);
  \draw[-] (1,-1) to (1.8,-1);
  \node at (2.2,0) {$\vec k_1$};
  \node at (2.2,-1) {$\vec k_2$};
  \end{feynman}
\end{tikzpicture}}}

\def\LCtrianglect{\scalebox{1.}{\begin{tikzpicture}[baseline={([yshift=-2.5]current bounding box.center)}]
  \begin{feynman}
    \node at (1,0.3) {\footnotesize \phantom{$\vec{k_1'}$}};
    \node at (1,-1.3) {\footnotesize \phantom{$\vec{k_2'}$}};
    \node at (-0.05,-0.2) {$\vec{k}$};
  \draw[-](-0.15,-0.5) to (0.35,-0.5);
  \draw(0.5,-0.5) circle [radius=0.15];
  \draw(0.6,-0.6) to (0.4,-0.4);
  \draw(0.6,-0.4) to (0.4,-0.6);
  \node at (0.8,-0.5) {\tiny 3};
  \draw(0.85,0) arc (90:180:0.35);
  \draw(0.85,-1) arc (270:180:0.35);
  \draw[-] (0.85,0) to (1.8,0.0);
  \draw[-] (0.85,-1) to (1.8,-1);
  \node at (2.2,0) {$\vec k_1$};
  \node at (2.2,-1) {$\vec k_2$};
  \end{feynman}
\end{tikzpicture}}}

\def\oneloop{\scalebox{1.}{\begin{tikzpicture}[baseline={([yshift=-2.5]current bounding box.center)}]
  \begin{feynman}
  \draw(1,0) circle [radius=0.5];
  \end{feynman}
\end{tikzpicture}}}

\def\ctmass#1{\scalebox{1.}{\begin{tikzpicture}[baseline={([yshift=-2.5]current bounding box.center)}]
  \begin{feynman}
  \draw(0,0) circle [radius=0.15];
  \draw(-0.1,-0.1) to (0.1,0.1);
    \draw(-0.1,0.1) to (0.1,-0.1);
  \node at (0, 0.3) {\tiny #1};
  \node at (0,-0.3) {\phantom{\tiny #1}};
  \end{feynman}
\end{tikzpicture}}}

\def\twoloopsct{\scalebox{1.}{\begin{tikzpicture}[baseline={([yshift=-2.5]current bounding box.center)}]
  \begin{feynman}
    \draw(-0.5,-0.5) to (0.,-0.5);
    \draw(0.5,0) arc (90:270:0.5);
    \draw(0.65,0.) circle [radius=0.15];
    \draw(0.65-0.1,0.-0.1) to (0.65+0.1,0.+0.1);
    \draw(0.65-0.1,0.+0.1) to (0.65+0.1,0.-0.1);
    \node at (0.65, 0.+0.3) {\tiny 2};
    \draw[color=white](0.65,-1.5) [radius=0.15];
    \draw(0.8,0) arc (90:-90:0.5);
    \draw(0.5,-1) to (0.8,-1);
    \draw(1.3,-0.5) to (1.8,-0.5);
    \end{feynman}
\end{tikzpicture}}}

\def\onetotwoct{\scalebox{1.}{\begin{tikzpicture}[baseline={([yshift=-2.5]current bounding box.center)}]
  \begin{feynman}
    \draw(-0.5,-0.5) to (0.,-0.5);
    \draw(0.5,0) arc (90:270:0.5);
    \draw(0.65,0.) circle [radius=0.15];
    \draw(0.65-0.1,0.-0.1) to (0.65+0.1,0.+0.1);
    \draw(0.65-0.1,0.+0.1) to (0.65+0.1,0.-0.1);
    \node at (0.65, 0.+0.3) {\tiny 2};
    \draw[color=white](0.65,-1.5) [radius=0.15];
    \draw(0.8,0) to (1.3,0);
    \draw(0.5,-1) to (1.3,-1);
    \end{feynman}
\end{tikzpicture}}}

\def\twoloopsint{\scalebox{1.}{\begin{tikzpicture}[baseline={([yshift=-2.5]current bounding box.center)}]
  \begin{feynman}
  \draw(1,0) circle [radius=0.5];
  \draw[-] (0.6464,0.3535) to (1.35,-0.35);
  \end{feynman}
\end{tikzpicture}}}

\def\halfcircle{\scalebox{1.}{\begin{tikzpicture}[baseline={([yshift=-2.5]current bounding box.center)}]
  \begin{feynman}
  \draw(1,0) arc (90:270:0.5);
  \draw[-] (1,0) to (1.4,0);
  \draw[-] (1,-1) to (1.4,-1);
  \end{feynman}
\end{tikzpicture}}}

\def\halfcircleandloop{\scalebox{1.}{\begin{tikzpicture}[baseline={([yshift=-2.5]current bounding box.center)}]
  \begin{feynman}
  \draw(1,0) arc (90:270:0.5);
  \draw[-] (1,0) to (1.5,-1);
  \draw[-] (1,0) to (1.8,0.0);
  \draw[-] (1,-1) to (1.8,-1);
  \end{feynman}
\end{tikzpicture}}}

%% file: biblio.bbl
%